\documentclass[a4paper,11pt]{article}

\pdfoutput=1 
\usepackage{jheppub} 

\usepackage{bbm}
\usepackage{multirow}

\usepackage{float}

\newcommand{\mr}{\mathrm}
\renewcommand{\arraystretch}{1.0}
\newcommand{\refeq}[1]{(\ref{#1})}

\newcommand\sfrac[2]{{\textstyle \frac{#1}{#2}}}

\title{Seesaw neutrino masses with a second Higgs doublet added}
\author[a]{D. Jur\v{c}iukonis,}
\author[a,b]{T. Gajdosik,}
\author[a]{and A. Juodagalvis}

\affiliation[a]{Vilnius University, Institute of Theoretical Physics and Astronomy, \\ A.\ Go\v{s}tauto st.\ 12, Vilnius 01108, Lithuania}
\affiliation[b]{Vilnius University, Physics Faculty, \\ Saul\.{e}tekio al.\ 9, Vilnius 10222 , Lithuania}

\emailAdd{darius.jurciukonis@tfai.vu.lt}

\abstract{
We study parameters of an extended standard model. The neutrino sector
is enlarged by one or two right-handed singlet fields and the Higgs
sector contains one additional doublet.
One-loop radiative corrections generate the mass for the light neutrino
fields.
The numerical analysis is performed
varying the masses of heavy neutrinos and
of the additional neutral Higgses.
The parameters of the neutrino sector,
allowing for the seesaw type-I mechanism,
are restricted by experimental
neutrino oscillation data.
Both normal and inverted hierarchies of the light neutrino
masses are discussed.
}

\begin{document} 
\maketitle
\flushbottom

\section{Introduction}

The precise interpretation of the neutral lepton fields in the particle
physics Lagrangian is not settled yet, owing to the very small mass of
the known neutrinos and the weakness of their interaction with other
particles~\cite{PDG2014}. The observed neutrino oscillations support the notion that
neutrinos have non-vanishing masses, calling for a modification of the
Standard Model (SM).
The size of the neutrino mass is not the only puzzle to solve.
Absence of an electrical charge allows neutrinos to be
their own antiparticles. The nature of the neutrinos -- whether they
are Dirac or Majorana particles -- might be determined by future
experiments.

The Standard Model considers neutrinos as massless.
Adding heavy right-handed neutral
singlets and additional Higgs doublets, the authors of
ref.~\cite{Grimus:1989pu} combined the seesaw mechanism (type-I)
with the radiative mass generation. The spontaneous
symmetry breaking (SSB) of the SM gauge group leads to a
Dirac mass term for neutrinos. The assumption that neutrinos are
Majorana particles allows an additional term in the Lagrangian,
namely, the Majorana mass term for the heavy singlets.

The model parameters allow small masses of the light
neutrinos that are compatible with the experimental observations. We
use this model in the formulation of Grimus and
Lavoura~\cite{Grimus:2002nk,Grimus:2002prd}, restricting the number of
additional Higgs doublets to one. We consider only 1-loop corrections
to the neutrino mass matrix.  The case of three additional heavy
neutrinos added to three light neutrinos was studied e.g.\ in
ref.~\cite{AristizabalSierra:2011mn,Dev:2012sg}.  We assume
either one or two heavy neutrinos.  Our preliminary results were
presented at several
conferences~\cite{Jurciukonis:2012jz,Jurciukonis:2012ft,
  Gajdosik:2013gpa,Jurciukonis:2014sma,Gajdosik:2015jja}. Here we provide a
more complete description of the performed numerical analysis.

Our extended model has several subsets of the parameters. The neutrino
sector is characterized by the masses of the heavy neutrinos (either one
or two), and the strength of the coupling to the neutral Higgs fields.
The masses of three light neutrinos are the result of our model
parameters. They are subject to experimental constraints,
namely the experimental neutrino mass differences, $\Delta m^2_\odot$
and $\Delta m^2_\mathrm{atm}$,
as well as the experimental neutrino oscillation
angles $\theta_{12}$, $\theta_{13}$, and
$\theta_{23}$~\cite{Forero:2014bxa}.
To estimate the neutrino
oscillation angles from the neutrino mixing matrix we follow the
ideas of ref.~\cite{Xing:2011ur}. More details are given in
appendix~\ref{Appendix-oscillation-angles}.
It should be noted that experimental data is
usually interpreted in the ``$3\times3$'' neutrino mixing
model~\cite{PDG2014,Forero:2014bxa},
i.e.\ three flavoured neutrinos are considered
as mixed states of three neutrino mass-eigenstates. We did not
attempt to reinterpret the results in the context of an extended
neutrino model. We expect the effect to be negligible.

We parametrize the Higgs sector along the analysis
of~\cite{Haber:2010bw}.
The Yukawa couplings are parametrized similarly to Grimus and
Lavoura~\cite{Grimus:2002nk,Grimus:2002prd}, which coincide
with \cite{Haber:2010bw} in the Higgs sector.
For the numerical analysis we take the mass of
the SM Higgs boson as $m_{H_1^0}=125$~GeV~\cite{Aad:2015zhl}
and allow the masses of two heavier Higgs bosons
to vary in the range from 126 to 3000~GeV.

The outline of the paper is following. Section \ref{framework} reviews
the seesaw mechanism and the formalism of the two-Higgs-doublet model
as used in our analysis. Sections \ref{3x1} and \ref{3x2} describe
our main results, namely, the calculated light neutrino mass spectra
and the analysis of free parameters. Our findings are summarized in
section \ref{summary}. For completeness, the appendix
section \ref{Appendix-b-vectors} describes the features of the weight
vectors $b_i$ that relate the scalar Higgs fields to their mass
eigenfields, and section
\ref{Appendix-oscillation-angles} gives the details of the oscillation
angle calculation.

\section{Description of the model}
\label{framework}

We discuss an extension of the Standard Model with enlarged Higgs
and neutrino sectors. Our main interest is the neutrino sector.
Since we need the Higgs sector for the radiative neutrino masses,
we give a short overview of the properties of the Higgs sector
that we use in our calculations.

\subsection{The Higgs sector}
\label{framework:Higgs}
The authors of ref.~\cite{Haber:2006ue} discuss the basis independent
formulation of the general two-Higgs-doublet model (2HDM).
Using their definition of the Higgs basis, we can write the
two complex doublets of our model in a unique way
\begin{equation}\label{higgsbasis}
\phi_{1}
=
\left(
\begin{array}{c} G^{+} \\ \frac{1}{\sqrt{2}}
( v + \mathcal{H}^{0}_{1r} + i G^{0} )  \end{array}
\right)\ ,
\qquad
\phi_{2}
=
\left(
\begin{array}{c} \mathcal{H}^{+} \\ \frac{1}{\sqrt{2}}
( \mathcal{H}^0_{2r} + i \mathcal{H}^0_{2i} )  \end{array}
\right)
\enspace ,
\end{equation}
where the vacuum expectation value (vev) $v \simeq 246$~GeV and the
Goldstone bosons $G^{0}$ and $G^{+}$ appear only in the first
Higgs doublet $\phi_{1}$. The relations between the basis independent
parameters defining the Higgs potential and the parameters describing
the physical states are linear and can be easily inverted. This
feature allows us to use the vev, the masses of the physical Higgs
bosons, $m_{H_1^0}$, $m_{H_2^0}$, $m_{H_3^0}$, and $m_{H^+}$,
and their mixing angles $\alpha_{12}$ and $\alpha_{13}$ as input
parameters.

The mass eigenstate for the charged Higgs boson corresponds directly
to the field $\mathcal{H}^{+}$ with the mass $m_{H^+}$, but the
mass eigenstates for the neutral Higgs bosons with the masses
$m_{H_1^0}$, $m_{H_2^0}$, and $m_{H_3^0}$, respectively, are
linear superpositions of the neutral fields $\mathcal{H}^{0}_{1r}$,
$\mathcal{H}^0_{2r}$, and $\mathcal{H}^0_{2i}$. Following the
formulation of Grimus and Lavoura~\cite{Grimus:2002nk,Grimus:2002prd}
these linear superpositions are conveniently expressed by
\begin{equation}
\label{sumb}
h_k^0=
\phi_{b_k}^0=
 \sqrt{2} \, \mathrm{Re}(b_k^{\dagger} \overline{\phi}^0)
 =
\sqrt{2} \sum_{j=1}^{n_H} \mathrm{Re}(b_{kj}^* \overline{\phi}_j^0)
 =
 \frac{1}{\sqrt{2}} \sum_{j=1}^{n_{H}}
 \left( b^{*}_{kj} \overline{\phi}^{0}_j
      + b_{kj} \overline{\phi}^{0\,*}_j \right)
\enspace ,
\end{equation}
where $\overline{\phi}^{0}$ are the neutral parts of the Higgs doublets
without the vev: $\overline{\phi}_{1}^{0} = \phi_{1}^{0} - v/\sqrt{2}$
and $\overline{\phi}_{2}^{0} = \phi_{2}^{0}$.
The ''b-vectors'' are $2 n_H$ unit vectors $b_k \in
\mathbbm{C}^{n_H}$ of dimensions $n_H \times 1$. We discuss those
vectors in the general case in appendix~\ref{Appendix-b-vectors}.
There we also show how
to obtain the following parametric values for the vectors $b$:
\begin{equation}
\label{bVectSetTh}
b_{G^0} = \left( \begin{array}{c} i \\ 0 \end{array} \right)
, \hspace{0.2cm}
b_1 = \left( \begin{array}{c}
 \mathrm{c}_{12} \mathrm{c}_{13}
 \\ -\mathrm{s}_{12} - i \mathrm{c}_{12} \mathrm{s}_{13}
\end{array} \right)
, \hspace{0.2cm}
b_2 = \left( \begin{array}{c}
 \mathrm{s}_{12} \mathrm{c}_{13}
 \\ \mathrm{c}_{12} - i \mathrm{s}_{12} \mathrm{s}_{13}
\end{array} \right)
, \hspace{0.2cm}
b_3 = \left( \begin{array}{c}
 \mathrm{s}_{13} \\
 i \mathrm{c}_{13}
\end{array} \right)
\enspace ,
\end{equation}
where $\mathrm{c}_{1j} = \cos\alpha_{1j}$ and
$\mathrm{s}_{1j} = \sin\alpha_{1j}$ are given by the angles
that describe the mixing of the neutral Higgs fields.

Restricting ourselves to CP conserving cases we use the analysis
of ref.~\cite{Haber:2010bw}, where the authors discuss
the CP-invariant Higgs potential in the 2HDM framework
under various basis-independent conditions.
The possible overall phase, that can be written in front of
the second Higgs doublet and that acts like a mixing angle
$\alpha_{23}$ between $\mathcal{H}^0_{2r}$ and $\mathcal{H}^0_{2i}$,
is used to define the CP-property of the mass eigenstates,
corresponding to their coupling to fermions,
taking $H_{2}^{0}$ to be CP-even and $H_{3}^{0}$ to be CP-odd.
This justifies the disctinction between the fields $h_{k}^{0}$
and $H_{k}^{0}$.

Having a fixed SM Higgs mass $m_{H^0_1}$ and assuming it
to be smaller than the other two,
non-degenerate neutral Higgs boson masses, we have four conditions
(case~I, case~II, case~IIIa with $m_{H^0_2}<m_{H^0_3}$, and
case~IIIb with $m_{H^0_2}>m_{H^0_3}$), which are listed in
Table~\ref{table1}.
A study reported in~\cite{Haber:2006ue} suggests that generality is not
lost assuming $-\frac{\pi}{2} \leqslant \alpha_{12},\alpha_{13} <
\frac{\pi}{2}$. We performed the numerical analysis of the neutrino
mass spectrum considering the named cases. In
some situations we refer to those cases as ``scenarios.''

\def\BigColSep{\setlength{\arraycolsep}{0pt}}
\renewcommand{\arraystretch}{1.1}
\begin{table*}
\begin{center}
\begingroup\BigColSep
\begin{tabular}{|c|c|c|c|}
\hline \hline
 & I & II & III \\
\hline
 & $\alpha_{12} = 0$ & $\alpha_{13} = 0$
 & $ \begin{array}{c} \alpha_{12} = 0 \\\alpha_{13} = 0 \end{array}$
\\
\hline
 & $m_{H^0_2}<m_{H^0_3}$ &$m_{H^0_2}>m_{H^0_3}$ &
 $
  \begin{array}{cc}
  \mathrm{(a)}\ & m_{H^0_2}<m_{H^0_3} \\
  \mathrm{(b)}\ & m_{H^0_2}>m_{H^0_3}
  \end{array}
 $
\\
\hline
$b_1$
& $\left( \begin{array}{c} \mathrm{c}_{13} \\
  -i \mathrm{s}_{13} \end{array} \right)$
& $\left( \begin{array}{c} \mathrm{c}_{12} \\
  - \mathrm{s}_{12} \end{array} \right)$
& $\left( \begin{array}{c} 1 \\ 0 \end{array} \right)$
\\
\hline
$b_2$
& $\left( \begin{array}{c} 0 \\ 1 \end{array} \right)$
& $\left( \begin{array}{c} \mathrm{s}_{12} \\
  \mathrm{c}_{12} \end{array} \right)$
& $\left( \begin{array}{c} 0 \\ 1 \end{array} \right)$
\\
\hline
$b_3$
& $\left( \begin{array}{c} \mathrm{s}_{13} \\
  i \mathrm{c}_{13} \end{array} \right)$
& $\left( \begin{array}{c} 0 \\ i \end{array} \right)$
& $\left( \begin{array}{c} 0 \\ i \end{array} \right)$
\\
\hline\hline
\end{tabular}
\endgroup
\end{center}
\caption{Basis-independent conditions for a CP-conserving 2HDM scalar
 potential and vacuum~\cite{Haber:2010bw}. $\alpha_{ij}$ label the
 mixing angles of neutral Higgses, $ m_{H^0_2}$ and $ m_{H^0_3}$
 denote the masses for CP-even and CP-odd Higgses.} \label{table1}
\end{table*}
\renewcommand{\arraystretch}{1.0}

\subsection{The Yukawa couplings}
\label{framework:Yukawa}
Using the
vector-and-matrix notation, the Yukawa Lagrangian for the leptons is
expressed by~\cite{Grimus:2002nk,Grimus:2002prd}
\begin{equation}
\label{Yukawa}
\mathcal{L}_\mathrm{Y} = - \sum_{k=1}^{n_H=2}\,
\left( \phi_k^\dagger \bar \ell_R \Gamma_k
+ \tilde \phi_k^\dagger \bar \nu_R \Delta_k \right)
    \left(\begin{array}{c}
      \nu_L \\
      \ell_L
    \end{array}\right)
+ \mathrm{H.c.} ,
\end{equation}
where $\tilde \phi_k = i \tau_2 \phi_k^\ast$. The quantities
$\ell_R$ and $\nu_R$ are the vectors of the
right-handed charged leptons and the right-handed projection of
the neutrino singlets, respectively.
$\ell_L$ and $\nu_L$ form the lepton doublet under the
weak interactions
and combine with the Higgs doublets $\phi_{k}$ to form
$SU(2)_{\mathrm{weak}}$-invariant terms.
They are also vectors in generation-space, $n_L = 3$ denoting
the three generations of the SM.
The Yukawa
coupling matrices $\Gamma_k$ have a dimension $n_L \times n_L$, while
$\Delta_k$ have a dimension $n_R \times n_L$, where $n_R$ is
the number of singlet neutrino fields.

Taking the bilinear terms of eq.~(\ref{Yukawa}), which means taking
only the vev from the Higgs doublets, we get the Dirac mass terms
for charged leptons and neutrinos:
\begin{equation}\label{M_ell}
M_\ell = \frac{v}{\sqrt{2}}\, \Gamma_1
\enspace\doteq
\mathrm{diag} \left( m_e, m_\mu, m_\tau \right)
\end{equation}
and
\begin{equation}\label{M_D}
 M_D = \frac{v}{\sqrt{2}} \Delta_1
\enspace .
\end{equation}
These have to be diagonalized using a singular-value decomposition (SVD)
like in the SM to get the correct definition for the mass eigenstates
that will describe the physical particles. Having done this
transformation to the mass eigenstates, which we write down as the
fields appearing in eq.~(\ref{Yukawa}), the respective transformation
matrices reappear in two unique combinations, $V_{\mathrm{CKM}}$
and $V_{\mathrm{PMNS}}$, in the interactions
with the charged gauge bosons $W^{\mp}$ or the charged scalar bosons
$H^{+}$ and $G^{+}$, giving the charged current Lagrangian
\begin{equation}
\mathcal{L}_{\mathrm{cc}}
= \frac{g}{\sqrt{2}}\, W_\mu^-
\bar{\ell}_{L} \gamma^{\mu} P_L V_{\mathrm{PMNS}} \nu_{L} + \mathrm{H.c.}
\enspace ,
\end{equation}
where $g$ is the $SU(2)$ gauge coupling constant. We give this part of
the Lagrangian only as a reference, to show what neutrino experiments
measure, as this PMNS matrix $V_{\mathrm{PMNS}}$ is the basis for the
interpretation of experimental data in the ``$3\times3$'' neutrino mixing
model~\cite{PDG2014}.

\subsection{Neutrinos at tree level}
\label{framework:neutrinos}
The singlet neutrinos, added to the SM, are neutral with respect to all
gauge groups of the SM. This offers the possibility that they are
Majorana particles, allowing to write a Majorana mass term for them.
Since the Lagrangian has to be a scalar with respect to Lorentz
transformations, we have to combine a spinor with itself in a Lorentz
invariant way. The only chance for Dirac spinors is to use the charge
conjugation matrix $\mathbf{C}$, which also appears in the definition
of the Lorentz covariant conjugation\footnote{For a very clear and
exhaustive description of the difference between Majorana and Dirac
spinors, see ref.~\cite{Pal:2010ih}.} for spinors
\begin{equation}\label{lcc}
\hat{\psi}
:=
\gamma^{0} \mathbf{C} \psi^{*}
=
- \mathbf{C} \bar{\psi}^{\top}
\enspace .
\end{equation}
The Majorana condition can now be written as
\begin{equation}\label{majorana-condition}
\hat{\psi}_{M}
=
\eta_{\psi} \psi_{M}
\enspace ,
\end{equation}
where $\eta_{\psi}$ is the Majorana phase. Assuming $\nu_{R}$ to
be $n_{R}$ Majorana fermions we can write down a Majorana mass term as
\begin{equation}\label{majorana-mass-term}
\mathcal{L}_{\mathrm{Majorana-mass}}
=
- \sfrac{1}{2} \bar{\nu}_{R} M_{R} \hat{\nu}_{R} + H.c.
=
  \sfrac{1}{2} \bar{\nu}_{R} M_{R} \mathbf{C} \bar{\nu}_{R}^{\top}  + H.c.
\enspace ,
\end{equation}
where the order of $M_{R}$ and $\mathbf{C}$ is irrelevant, as these
matrices act on different indices of the spinor $\nu_{R}$: $\mathbf{C}$
is a $4\times4$-Dirac matrix, connecting the spinor indices of $\nu_{R}$,
whereas $M_{R}$ is a symmetric $n_{R}\times n_{R}$ matrix, acting on the
''generation'' index of $\nu_{R}$.
Since the mechanism generating the Majorana mass is not known,
we assume the singlets $\nu_{R}$ to be
already in the mass eigenstate of $M_{R}$. This means, we
assume $M_{R}$ to be diagonal, containing the Majorana masses of the
heavy singlets: $M_{R} = \hat{M}_{R}$.

Together with the Dirac mass, coming from the Yukawa terms
eq.~(\ref{Yukawa}), the mass terms for the neutrinos are
\begin{eqnarray}\label{nu-mass}
  \mathcal{L}_{\nu-\mathrm{mass}}
&=&
- \bar{\nu}_{R} M_{D} \nu_{L}
- \sfrac{1}{2} \bar{\nu}_{R} M_{R} \hat{\nu}_{R}
+ H.c.
\nonumber\\
&=&
- \sfrac{1}{2} \bar{\nu}_{R} M_{D} \nu_{L}
- \sfrac{1}{2} \bar{\hat{\nu}}_{L} M_{D}^{\top} \hat{\nu}_{R}
+ \sfrac{1}{2} \bar{\nu}_{R} M_{R} \mathbf{C} \bar{\nu}_{R}^{\top}
+ H.c.
\nonumber\\
&=&
- \sfrac{1}{2}
\left(\begin{array}{cc}
  \bar{\hat{\nu}}_{L} & \bar{\nu}_{R}
\end{array}\right)
\left(\begin{array}{cc}
  0 & M_{D}^{\top} \\
  M_{D} & \hat{M}_{R}
\end{array}\right)
\left(\begin{array}{c}
  \nu_{L}
 \\ \hat{\nu}_{R}
\end{array}\right)
+ H.c.
\end{eqnarray}
and can be written in a compact form by introducing an
$(n_L+n_R) \times (n_L+n_R)$ symmetric neutrino mass matrix
\begin{equation}
\label{Mneutr}
\renewcommand{\arraystretch}{0.8}
M_{\nu} =
\left(\begin{array}{cc} 0 & M_D^{\top} \\
M_D & \hat{M}_R \end{array} \right)
\enspace .
\end{equation}
The neutrino mixing matrix $M_{\nu}$ can be
diagonalized~\cite{Grimus:1989pu,Grimus:2002prd} using the properties
of the singular-value decomposition of a symmetric matrix
\begin{equation}
\label{Mtotal}
U^{\top} M_{\nu}\, U = \hat m
= \mathrm{diag}
\left( m_{l_1}, m_{l_2}, m_{l_3}, m_{h_1}, \dots , m_{h_{n_{R}}} \right),
\end{equation}
where $m_{l_i}$ and $m_{h_i}$ are real and non-negative with
mass-ordering $m_{l_1} \leq m_{l_2} \leq m_{l_3}$ and $m_{h_1}\leq
\dots \leq m_{h_{n_{R}}}$.
In order to implement the seesaw mechanism
\cite{GellMann:1980vs,Schechter:1980gr} we assume that the elements of
$M_D$ are of order $m_D$ and those of $M_R$ are of order $m_R$, with
$m_D \ll m_R$. Then, the neutrino masses $m_{l_i}$ with
$i=1,\ldots,n_L$ (where $n_L=3$), are of order $m_D^2/m_R$, while the
masses $m_{h_i}$ with $i = 1,n_R$ (where $n_R=1$ or $2$), are of order $m_R$.
It is useful to
decompose the $(n_L+n_R) \times (n_L+n_R)$ unitary matrix $U$ as
\cite{Grimus:1989pu,Grimus:2002prd}
\begin{equation}\label{U}
U = \left( \begin{array}{c} U_L \\ U_R^\ast \end{array} \right),
\end{equation}
where the submatrix $U_L$ is $n_L \times (n_L+n_R)$ and the submatrix
$U_R$ is $n_R \times (n_L+n_R)$. These submatrices obey certain
unitarity relations:
\begin{equation}\label{uni}
U_L U_L^\dagger = \mathbbm{1}_{n_L} \,, \quad
U_R U_R^\dagger = \mathbbm{1}_{n_R} \,, \quad
U_L U_R^{\top} = 0_{n_L \times n_R} \, \quad \mathrm{and} \quad
U_L^\dagger U_L + U_R^{\top} U_R^* = \mathbbm{1}_{n_L+n_R}
\enspace .
\end{equation}
Combining with eq.~(\ref{Mtotal}), we can obtain the following relations:
\begin{equation}\label{Urelat}
U_L^* \hat{m} U_L^\dagger = 0, \quad
U_R \hat{m} U_L^\dagger = M_D \quad \mathrm{and} \quad
U_R \hat{m} U_R^{\top} = M_R
\enspace .
\end{equation}

With these submatrices of $U$,
the left- and right-handed neutrinos
can be written as linear superpositions
of the $n_L+n_R$ physical Majorana neutrino fields $\chi_i$:
\begin{equation}\label{chi}
\nu_L = U_L P_L \chi
\quad \mathrm{and} \quad
\hat{\nu}_R = U_R^{*} P_L \chi
\quad \mathrm{or} \quad
\nu_R = U_R P_R \chi
\enspace ,
\end{equation}
where $P_L$ and $P_R$ are the projectors of chirality.

Switching to the physical Majorana mass states $\chi$, we have to
express the couplings using the matrices $U_{L}$ and $U_{R}$.
The loop corrections, described in the next subsection, depend on
the neutrino couplings to the $Z$-boson and to the neutral Higgses.
Interaction with the $Z$ boson is given by
\begin{equation}
\mathcal{L}_{\mathrm{nc}}^{(\nu)} = \frac{g}{4 c_w}\, Z_\mu
\bar \chi \gamma^\mu \left[ P_L \left( U_L^\dagger U_L \right)
- P_R \left( U_L^{\top} U_L^\ast \right) \right] \chi\, ,
\label{Zinter}
\end{equation}
where $c_w$ is the cosine of the Weinberg angle.
The Yukawa couplings for the neutral scalars
take the form
\begin{align}
\mathcal{L}_\mathrm{Y}^{(\nu)} \left( h_{k}^0 \right) = &
 -\frac{1}{2 \sqrt{2}}\,
 \sum\limits_{k=1}^{2n_H=4} h_{k}^0 \, \bar \chi \Big[
\left( U_R^\dagger \Delta_{b_k} U_L
+ U_L^{\top} \Delta_{b_k}^{\top} U_R^\ast \right) P_L \notag \\
&+ \left( U_L^\dagger \Delta_{b_k}^\dagger U_R
+ U_R^{\top} \Delta_{b_k}^\ast U_L^\ast \right)P_R
\Big] \chi \, ,
\label{neutralYuk}
\end{align}
where we treat the Goldstone boson $G^{0}$ as $h_{4}^{0}$.
The Yukawa coupling $\Delta_{b_k}$ is the result of rewriting
the Yukawa Lagrangian eq.~(\ref{Yukawa}) using the physical Higgs
fields defined in eq.~(\ref{sumb}):
\begin{equation}
\Delta_{b_k} = \sum_{j=1}^{n_{H}} (b_k)_{j} \Delta_{j} .
\end{equation}

\subsection{Loop corrections to the neutrino masses}
\label{framework:loops}
We are interested in radiatively generated neutrino masses at one-loop
level. The largest influence from the corrections to the neutrino
mass matrix has the neutrino Majorana mass term $\delta M_L$,
since this submatrix is zero at tree level,
$\left.M_L\right|_{\mathrm{tree}}=0$.
The contributions from
charge-changing currents are
subdominant~\cite{Grimus:2002nk,Grimus:2002prd}.

We calculate the radiative light neutrino masses following
ref.~\cite{Grimus:2002nk}. Once the one-loop corrections are
taken into account the neutral fermion mass matrix is given by
\begin{equation}\label{M1}
M^{(1)}_\nu = \left( \begin{array}{cc}
\delta M_L & M_D^{\top}+\delta M_D^{\top} \\
M_D+\delta M_D  & \hat{M}_R+\delta M_R\end{array} \right)\approx
\left( \begin{array}{cc}
\delta M_L & M_D^{\top} \\
M_D  & \hat{M}_R\end{array} \right) ,
\end{equation}
where the $0_{3\times3}$ matrix appearing at tree level (\ref{Mneutr})
is replaced by a symmetric matrix $\delta M_L$. This correction
dominates among all the sub-matrices of corrections. The one-loop
corrections to $\delta M_L$ originate via the self-energy function
$\Sigma_L^{S(X)}(0)$ (where $X=Z,G^0,H_b^0$) that arises from the
self-energy Feynman diagrams. The contributions
$\Sigma_L^{S}(p^2)$ are evaluated at zero external
momentum squared ($p^2=0$). The neutrino couplings to
the $Z$, Higgs $H_b^0$ and Goldstone $G^0$ bosons
are determined by eqs.~(\ref{Zinter}) and (\ref{neutralYuk}). Each
diagram contains a divergent piece but the sum of the three
contributions yields a finite result. The expression for one-loop
corrections is given by (see e.g.\ \cite{Grimus:2002nk})
\begin{eqnarray}
\delta M_L &=&
\sum_{k=1}^{3} \frac{1}{32 \pi^2}\, \Delta_{b_k}^{\top} U_R^\ast \hat m
\left(
  \frac{\hat {m}^2}{m_{H^0_k}^2}-\mathbbm{1}
 \right)^{-1}\hspace{-5pt}
 \ln\left(\frac{\hat {m}^2}{m_{H^0_k}^2}\right) U_R^\dagger \Delta_{b_k} \notag \\
&&+ \frac{3 g^2}{64 \pi^2 m_W^2}\, M_D^{\top} U_R^\ast \hat m
\left(
  \frac{\hat {m}^2}{m_Z^2}-\mathbbm{1}
 \right)^{-1}\hspace{-5pt}
 \ln\left(\frac{\hat {m}^2}{m_Z^2}\right) U_R^\dagger M_D,
\label{corrections}
\end{eqnarray}
where the sum index $k$ runs over all neutral physical Higgses $H^0_k$.

\section{Results for the case $\mathbf{n_R=1}$}
\label{3x1}

\subsection{General parameterization}

First we consider a minimal extension of the standard model by adding
only one right-handed neutrino field $\nu_R$ to three left-handed
fields $\nu_L$. This simple model is useful because it has
a small number of parameters. It allows us to obtain
relations between the free parameters and the light neutrino masses.

For the study using the general parameterization
(reported in \cite{Jurciukonis:2012jz}), we fix the Higgs
sector by choosing a specific CP-conserving set of vectors~$b$:
\begin{equation}\label{specB}
b_{G^0}=\left(\begin{array}{c} i\\0 \end{array}\right),\quad
b_1 = \left(\begin{array}{c} 1\\0 \end{array}\right),\quad
b_2 = \left(\begin{array}{c} 0\\i \end{array}\right),\quad
b_3 = \left(\begin{array}{c} 0\\1 \end{array}\right).
\end{equation}

We use a parameterization of the Yukawa matrices
$\Delta_1$ and $\Delta_2$ in the following form:
\begin{equation}
\Delta_i = \frac{\sqrt{2}\, m_D}{v}\,\vec{a}_i^{\top} ,\ \ i=1,2.
\end{equation}
We further assume that the linearly independent vectors are
normalized, $|\vec{a}_1|=|\vec{a}_2|=1$, and
equal total strength of the couplings.
Using the block form,
diagonalization of the symmetric neutrino mass matrix at tree level
$M_{\nu}^{(0)}$ (\ref{Mneutr}) can be written as
\begin{equation}\label{Mneutr1}
U_{\mathrm{tree}}^{T}M_{\nu}^{(0)}U_{\mathrm{tree}} =
U_{\mathrm{tree}}^{T}\left(
 \begin{array}{cc}
   0_{3 \times 3} & m_D \vec{a}_1 \\
   m_D \vec{a}_1^{\top} & \hat{M}_R
 \end{array}
\right)U_{\mathrm{tree}}=
\left(
 \begin{array}{cc}
  \hat{M}_l^{(0)} & 0 \\
  0 & \hat{M}_h^{(0)}
 \end{array}
\right).
\end{equation}
The non-zero masses in $\hat{M}_l^{(0)}$ and $\hat{M}_h^{(0)}$ can be
determined analytically by finding the eigenvalues of the hermitian
matrix $M_{\nu}^{(0)}M^{(0)\dagger}_{\nu}$. These eigenvalues are squares
of the neutrino masses, $\hat{M}_l^{(0)}=\text{diag}(0,0,m_l^{(0)})$ and
$\hat{M}_h^{(0)}=m_h^{(0)}$. The solutions
\begin{align}
m^2_D=&m_h^{(0)}m_l^{(0)}\ , \label{mD}\\
m^2_R=&\left(m_h^{(0)}-m_l^{(0)}\right)^2 \approx
\left(m_h^{(0)}\right)^2 \label{mR}
\end{align}
correspond to the seesaw mechanism.

We diagonalize the tree-level neutrino mass matrix $M_{\nu}^{(0)}$
using a diagonalization matrix $U_{\mathrm{tree}}$ made of two
diagonal matrices of phases and three rotation matrices:
\begin{equation}
U_{\mathrm{tree}}=\hat{U}_{\phi}(\phi_i)U_{12}(\alpha_1)
  U_{23}(\alpha_2)U_{34}(\beta)\hat{U}_i,
\end{equation}
The angle $\beta$ is determined by the masses $m_l^{(0)}$ and
$m_h^{(0)}$: $\tan^2(\beta)=m_l^{(0)}/m_h^{(0)}$. The dependency of
$\phi_i$ and $\alpha_i$ on $m_D$ and ${\vec a}_1$ may be expressed
analytically. The diagonalization matrix $U_{\mathrm{tree}}$ is
unitary because the rotation matrices $U_{ij}$ and the diagonal phase
matrices $\hat{U}_{\phi}$ and $\hat{U}_i$ are all unitary.

Diagonalization of the neutrino mass matrix with the one-loop
corrections included, $M_{\nu}^{(1)}$ in eq.~\refeq{M1},
is performed numerically using a unitary matrix
\begin{equation}
U_{\text{loop}}=U_{\text{egv}}\hat{U}_{\varphi}(\varphi_1,\varphi_2,\varphi_3),
\end{equation}
where $U_{\text{egv}}$ is an eigenmatrix of
$M^{(1)\dagger}_{\nu}M^{(1)}_{\nu}$, and
$\hat{U}_{\varphi}$ is a phase absorption matrix. As discussed already
in ref.~\cite{Grimus:1989pu}, the heaviest light neutrino obtains mass
at tree level from the seesaw mechanism. The second light neutrino
obtains mass from radiative corrections. The lightest neutrino
remains massless.

The numerical evaluation of the model parameters and of the masses of the
light neutrinos proceeds in several steps. First, the mass
matrix for the tree level is constructed. The lightest neutrino
remains massless in this model, $m_{l_1}=0$. Using the central values
of the experimental neutrino mass differences, we estimate the mass of
the heaviest light neutrino and use it as an input parameter
$m_l^{\mathrm{in}}$. The value of $m_R$ is another input parameter.
Using the seesaw relations~\refeq{mD} and \refeq{mR}, we evaluate
$m_D=\sqrt{m_Rm_l^{\mathrm{in}}}$. The vector $\vec{a}_1$ is generated
randomly. This fully determines the tree-level neutrino mass matrix
$M_\nu^{(0)}$.
Solving the eigenvalue equation \refeq{Mneutr1} we obtain the
tree-level neutrino masses $m_l^{(0)}$, $m_h^{(0)}$, and the
diagonalization matrix $U_\mathrm{tree}$.
The next step is to evaluate the one-loop corrections to the
neutrino mass matrix $\delta M_L$. Additional input parameters of
$\vec{a}_2$ enter the procedure.
Diagonalization of the corrected
neutrino mass matrix $M_{\nu}^{(1)}$ yields masses for two light
neutrinos. If the obtained mass differences are compatible with the
experimental data on neutrino oscillations, the model parameter set is
kept. Otherwise, another set of input parameters
$\{m_R,\vec{a}_1,\vec{a}_2,m_{H_{2,3}^0}\}$ is generated.

The study suggested a lower limit of
830~GeV for the mass of the heavy singlet \cite{Jurciukonis:2012jz}.


\subsection{Reduced parameterization}

For the Higgs sector we use the values of the orthogonal complex
vectors $b$ listed in Table~\ref{table1}. The mass of the lightest
neutral Higgs is fixed at $m_{H^0_1}=125$~GeV. The masses of heavier
neutral Higgses $m_{H^0_2}$ and $m_{H^0_3}$ are generated randomly in
the range from 126 to 3000~GeV.

The light neutrino fields can be transformed in such a way
that $\vec{a}_1^{\top} = \bigl(0,0,1\bigr)$, and $\vec{a}_2^{\top} =
\bigl(0,n,n^{\prime}\bigr)$ with real numbers $m_D, n >0$, and a
complex number $n^{\prime}$. Due to the assumed normalisation
conditions $|\vec{a}_{1}|=|\vec{a}_{2}|=1$, there are only two
independent parameters, namely, the real number $n$ ($n\leq1$),
and a complex phase $\phi$:
\begin{align}
\vec{a}_1^{\top} =& \bigl(0,0,1\bigr)\ , \\
\vec{a}_2^{\top} =& \bigl(0,n,e^{i \phi}\sqrt{1-n^2}\bigr)\ .
\end{align}
A similar case was studied in sect.~4 of ref.~\cite{Grimus:1989pu}
without the assumption $|\vec{a}_2|=1$. Since the goal was to
demonstrate that one of the massless neutrinos can obtain mass through
the 1-loop radiative corrections, only the
expressions are given there. In addition, the masses of the lightest
and the heaviest light neutrinos do not change going from the tree
level to the 1-loop-corrected level
in the analysis of ref.~\cite{Grimus:1989pu}.

The numerical evaluation of the model parameters and the masses of the
light neutrinos in the case of the reduced parameterization is done
similarly to the general case, described above.
At the one-loop accuracy, our model will predict a vanishing mass
for the lightest neutrino. If we use this value directly,
the measured neutrino mass differences would lead to
highly restricted values of the neutrino masses.
In order to reduce the impact of a starting point to our
analysis we allow the lightest neutrino to have
a small non-vanishing mass, $m_{l_1}^{\mathrm{in}}$. Using
the central values of the experimental neutrino mass differences,
we determine the largest initial value of the light neutrino masses
$m_l^{\mathrm{in}}$.
Selecting the value of $m_R$, the value of $m_D$ is
evaluated from the seesaw relation, eq.~\refeq{mD}.
Since the vector $\vec{a}_1$ is fixed in this
case, the tree-level neutrino mass matrix $M_\nu^{(0)}$ is fully
defined as an input quantity at this point.

Solving the eigenvalue equation, we get the tree-level masses of
the heaviest light neutrino $m_l^{(0)}$ and of the heavy neutrino
$m_h^{(0)}$.
The diagonalization matrix
$U_{\mathrm{tree}}$ is constructed from a rotation matrix and a
diagonal matrix of phases: $U_{\mathrm{tree}}=U_{34}(\beta)\hat{U}_i$.
Then one-loop corrections to the neutrino mass matrix are
evaluated, and the parameters defining $\vec{a}_2$ enter into the further
evaluation.  Diagonalization of the 1-loop neutrino mass matrix
$M_{\nu}^{(1)}$ is performed numerically with a unitary matrix
$U_{\mathrm{loop}}$ as in the general parameterization case. After
this procedure two light neutrinos have masses $m_{l_2}^{(1)}$ and
$m_{l_3}^{(1)}$.
For the calculation of $\Delta m_{ij}^{2}$
we assume the mass of the lightest
neutrino to be the input value $m_{l_1}^{\mathrm{in}}$ with the
justification, that it could be generated by a two-loop contribution.
If the calculated mass differences
\begin{equation}
 \Delta \big(m_{12}^{(1)}\big)^2 =
  \big( m_{l_2}^{(1)} \big)^{2} -
  \big( m_{l_1}^{\mathrm{in}}\big)^{2}\quad
    \text{and}\quad
 \Delta \big(m_{23}^{(1)}\big)^{2} =
   \big( m_{l_3}^{(1)} \big)^{2} -
   \big( m_{l_2}^{(1)} \big)^{2},
\end{equation}
and the determined value of $\theta_{23}\approx\theta_{\mathrm{atm}}$
(as described in the next paragraph) are
compatible with the experimental data on oscillations,
the model parameter set
is kept. Otherwise, another set of input parameters
$\{m_{\ell_1}^{\mathrm{in}}, m_R, n, \phi, \alpha_{ij}, m_{H_{2,3}^0}\}$
is generated.

The results of the 1-loop corrected calculations are subject
to the constraints from the experimental data on solar and atmospheric
neutrino oscillations~\cite{Forero:2014bxa}. The neutrino mixing
angles are determined from a factorization of the diagonalization
matrix $U_{\mathrm{loop}}$ into terms where the PMNS matrix is
included explicitly, following the ideas of ref.~\cite{Xing:2011ur}.
(The method is described in detail in
appendix~\ref{Appendix-oscillation-angles}.) For the case $n_R=1$ it
is possible to find exact analytical expressions for the mixing
angles, Dirac and Majorana as well as non-physical phases. The
oscillation angles are constrained by the experimental data in the
sense that having a randomly generated set of the input parameters we
derive the 1-loop corrected results and require that the estimated
mixing angles are consistent with the experimental values in the
$3\sigma$ range. It should be noted that this reduced
parameterisation has only one non-vanishing neutrino mixing angle,
namely, the atmospheric neutrino mixing angle $\theta_{\mathrm{atm}}$,
because the vectors $\vec{a}_{1}$ and $\vec{a}_2$ have the same
vanishing component ($a_{11}=a_{21}=0$). In a more general case, all
three oscillation angles are non-zero.
\begin{figure}[t]
\begin{center}
\includegraphics[scale=0.95]{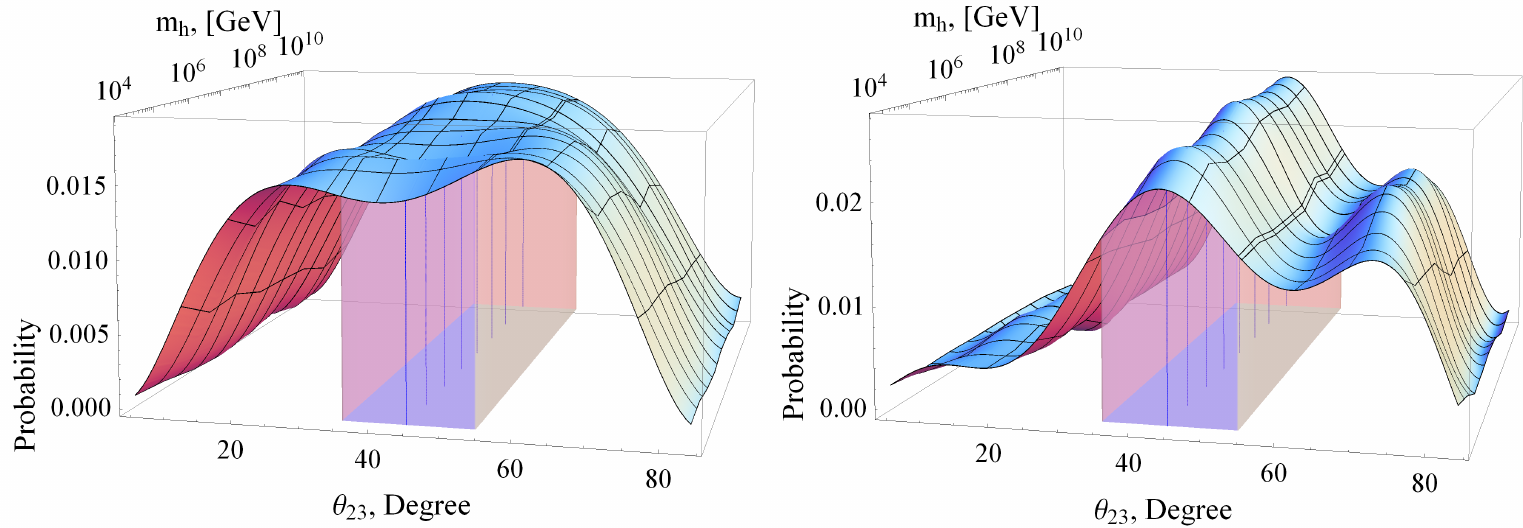}
\end{center}
\caption{(Color online) The 3D histograms of
 $\theta_{23}=\theta_{\mathrm{atm}}$ oscillation angles for $n_R=1$.
 The left plot represents the case II of the Table~\ref{table1}. The
 right plot shows the case I. The filled boxes indicate the $3\sigma$
 experimental boundaries \cite{Forero:2014bxa}, the blue vertical
 lines denote the experimental central value of $\theta_{23}$.}
 \label{picture1}
\end{figure}

Distributions of the atmospheric oscillation angles for the cases I
and II are shown in figure~\ref{picture1}. The plot on the right shows
the case I that is similar to the cases IIIa and IIIb. This
distribution is narrower than the one obtained in the case II.
The distribution for the case I has a well pronounced peak at around
$45$ degrees, while the
probability to have an oscillation angle $\theta_{23}$ in the range
from $25$ to $65$ degrees is nearly flat in the case II.

Since only two light neutrinos acquire mass in the case of $n_R=1$
with 1-loop correction included, only normal
ordering of neutrino masses is possible. Assuming that the lightest
neutrino mass $m_{l_1}=0$, we obtain a fixed spectrum of light
neutrinos with $m_{l_2}=8.7\pm0.3$~meV and $m_{l_3}=50.6\pm 2$~meV.
However, the peak of the derived distribution
of the oscillation angle $\theta_{\mathrm{atm}}$ is shifted from the
experimental value.
\begin{figure}[t]
\begin{center}
\includegraphics[scale=0.85]{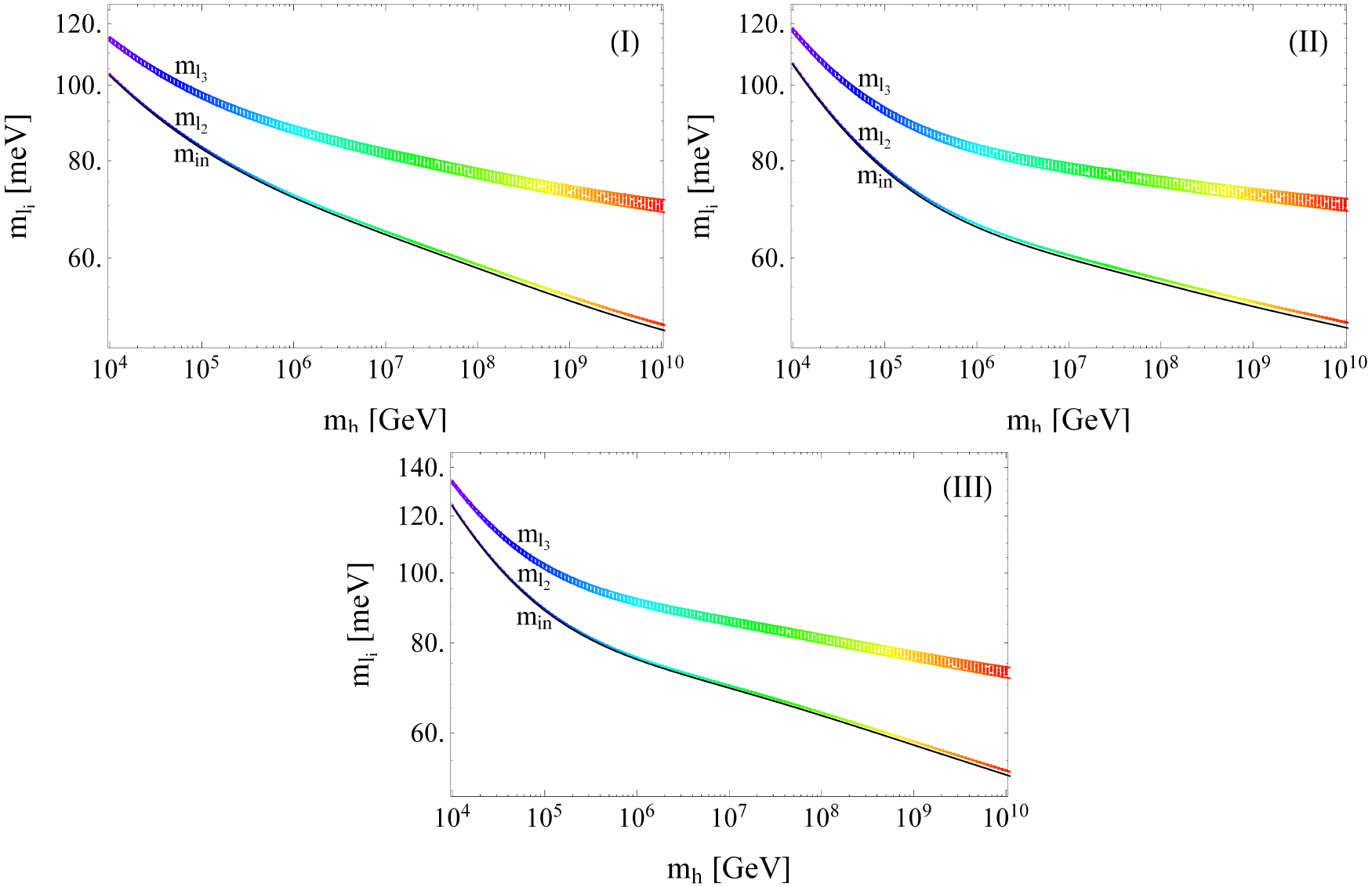}
\end{center}
\caption{(Color online) Calculated masses of two light neutrinos as a
 function of the heaviest neutrino mass $m_h$ for 3 different scenarios of
 Table~\ref{table1} and $n_R=1$. The black solid line represents the
 initial parameter $m_{\mr{in}}$. Values of $m_{l_2}$ and $m_{l_3}$
 are shown as a band where the color corresponds to the mass of the
 heavy neutrino $m_h$, given on the horizontal axis. This color
 code is also used in figures~\ref{picture3} and \ref{picture4}. }
\label{picture2}
\end{figure}

Figure~\ref{picture2} illustrates the neutrino mass spectrum for
three different scenarios given in Table~\ref{table1},
assuming $m_{l_1}=m_{\mr{in}}\neq0$.
All scenarios with $n_R=1$ produce
distributions of neutrino masses which differ from each other by value.
In the $n_R=2$ case (which is discussed below),
all distributions look very similar, regardless
of the scenario.
The cases IIIa and IIIb have similar distributions, therefore no
reference to the case 'a' or 'b' is given in the plot.
The mass of the heaviest neutrino $m_{l_3}$ reaches
the highest value of 140~meV, when $m_h=10^4$~GeV in the case III. The
lowest value of $m_{l_3}=70$~meV is obtained when $m_h=10^{10}$~GeV in
the cases I and II. The mass of the intermediate light neutrino
$m_{l_2}$ reaches the highest value of 121~meV, when $m_h=10^4$~GeV
in the case III, and the lowest value of $m_{l_2}=48$~meV is
reached when $m_h=10^{10}$~GeV in the case I.

We observe a particular relationship between the values of the masses of
light and heavy neutrinos. The calculated masses of the light
neutrinos decrease, if the heavy neutrino mass increases. This
dependence emerges from the relation of $m_{h}$ to
$\theta_{\mathrm{atm}}$. The value of $m_{\mr{in}}$ has to get lower
as $\nu_R$ gets heavier in order to keep the most probable value of
the oscillation angle within the experimental range of $3\sigma$ (this
range is marked by boxes in figure~\ref{picture1}). The narrow bands of
the values of $m_{l_2}$ and $m_{l_3}$ are formed by a relatively high
value of $m_{\mr{in}}$ and the restrictions on the mass
differences. The results suggest that the lowest limit of the heavy
neutrino mass is $10^4$~GeV.

\begin{figure}[t]
\begin{center}
\includegraphics[scale=0.97]{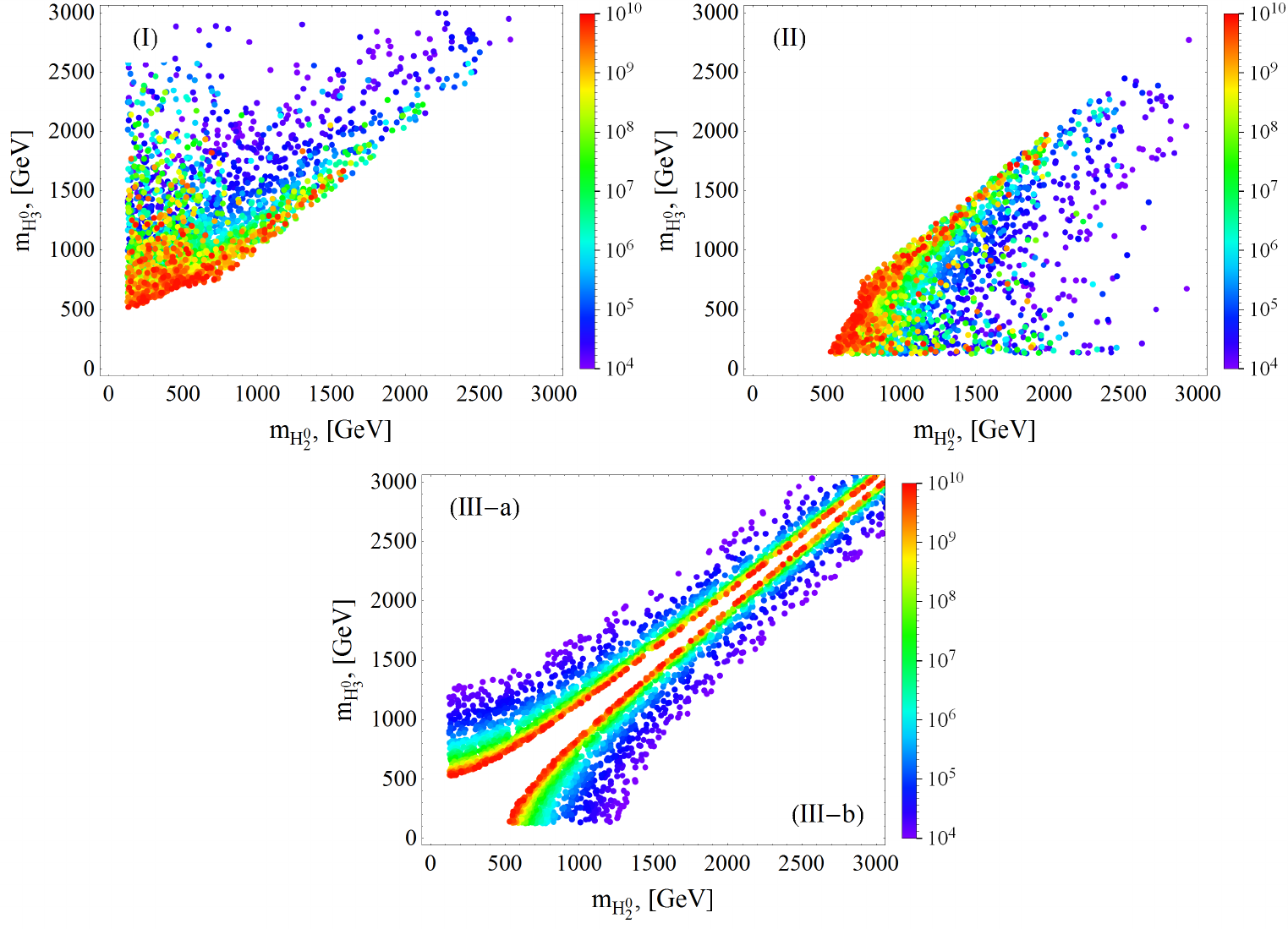}
\end{center}
\caption{(Color online) Values of the neutral Higgs masses
 $m_{H^0_2}$ and $m_{H^0_3}$ as a function of the heaviest
 neutrino mass $m_{h}$ for 4 different cases of
 Table~\ref{table1} in the $n_R=1$ model. The scale of the $m_h$
 values is shown on the right. The mass of the SM Higgs boson is
 fixed to $m_{H^0_1}=125$~GeV.}
\label{picture3}
\end{figure}

\begin{figure}[t]
\begin{center}
\includegraphics[scale=0.95]{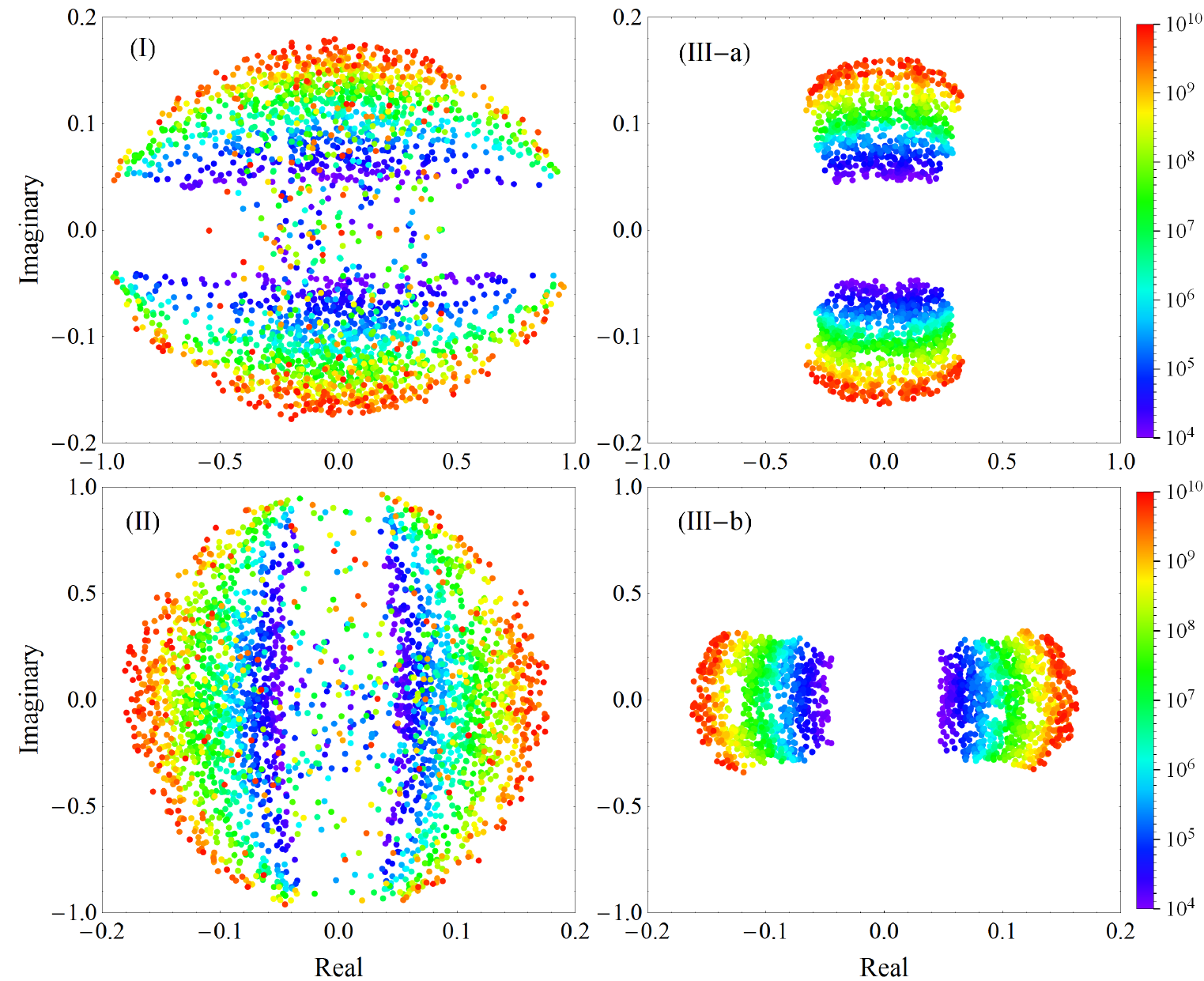}
\end{center}
\caption{(Color online) Values of the free parameter $a_{23}$
 as a function of the heaviest neutrino mass
 $m_{h}$ for the cases of Table~\ref{table1} and $n_R=1$.
 The scale of the $m_h$ values is shown on the
 right.
}
\label{picture4}
\end{figure}

The allowed values of the Higgs masses (other than the SM Higgs) are
illustrated in figure~\ref{picture3} as a function of the heavy singlet
mass $m_{h}$. A band structure is formed according to the choice of
vectors $b$ (see Table~\ref{table1}) and the values of the free
parameters $n$ and $\phi$, which are displayed in
figure~\ref{picture4}. When the mass of the heavy singlet $m_h$ is
increasing, the values of the
Higgs masses tend to decrease in the 2HDM model cases I and II.
This tendency is absent in the case III,
where the allowed values of $m_{H_{2,3}^0}$ form a narrow band.
Larger values of $m_h$ make the masses of
$m_{H_{2,3}^0}$ more similar, although their difference
does not disappear. The range of the allowed masses of the
heavier Higgs boson starts at 500~GeV.

The values of the Higgs masses displayed in figure~\ref{picture3}
satisfy the experimental restrictions~\cite{PDG2014} of the oblique
parameters S, T, and U, introduced by Peskin and Takeuchi
\cite{Peskin:1991sw}. These oblique parameters define combinations of
observables that quantify deviations from the SM
predictions. The experimental electroweak precision data yields values
compatible with SM.  We estimated these parameters using the algorithm
implemented in the two Higgs doublet model calculator
(2HDMC)~\cite{Eriksson:2009ws} taking recent values of SM
parameters~\cite{PDG2014}. The values of the oblique parameters are
functions of masses of the three neutral ($m_{H^0_1},$ $m_{H^0_2},$ and
$m_{H^0_3}$) and one charged ($m_{H^{\pm}}$) Higgs boson, and an
additional factor that is related to the mixing angles of the neutral
Higgses.  This factor, defined as $\sin(\beta-\alpha)$ in
ref.~\cite{Eriksson:2009ws}, equals to $\cos(\alpha_{13})$,
$\cos(\alpha_{12})$, and $1$ in the scenarios I, II, and III of
Table~\ref{table1}, respectively. The estimated masses of the neutral
Higgses and their mixing angles in our model are discussed above.  The
mass of the charged Higgs $m_{H^{\pm}}$ is allowed to take values in
the range of 126--3000~GeV. This defines the parameter space to
estimate the values of S, T, and U.  The calculated curves of S and U
are mostly contained within the experimental bounds. The
experimental limits on T are narrower as compared to the calculated
distribution of T.
A similar observation on the restrictive power
of the oblique parameters is made in ref.~\cite{Funk:2011ad}.
These limits can be used to restrict the mass of
the charged Higgs boson. If the values of all three oblique parameters
are compatible with the experimental data and the allowed value of
$m_{H^{\pm}}$ is within the studied range,
the tested set of model parameters
is kept. Otherwise, another set of parameters is
generated. Using a stronger constraint of $U=0$, which is also
compatible with SM, about 35\%--45\% of the values in
figure~\ref{picture3} satisfy the experimental
restrictions~\cite{PDG2014}.  In this case the majority of
$m_{H^0_{2,3}}$ values fall in the range of 126--1500~GeV.

Distributions of the free parameter $a_{23}$, part of the neutrino
couplings to the second Higgs doublet, are shown in
Fig.~\ref{picture4}. When the mass of the heavy singlet $m_h$ 
increases, the absolute value of the parameter $|a_{23}|$ also
increases. In case~I the real part of $a_{23}$ varies between 
$-1$ and $1$ and the imaginary part varies from $-0.18$ to $0.18$.
In case~II real and imaginary parts behave like interchanged: the real 
part of $a_{23}$ varies between $-0.18$ and $0.18$ while imaginary part 
varies from $-1$ to $1$. In case~III-a the real part of $a_{23}$ varies 
between $-0.34$ and $0.34$ while the imaginary part is restricted to the
intervals $(0.04 , 0.17)$ and $(-0.17,-0.04)$. In case~III-b real and 
imaginary parts are exchanged like between cases~I and II: the real part 
of $a_{23}$ is restricted to the intervals $(0.04 , 0.17)$ and 
$(-0.17,-0.04)$ while the imaginary part varies between $-0.34$ and $0.34$.

In summary, tuning the value of $m_{\mr{in}}$, and restricting the
light neutrino mass differences to the experimental central values
within $1\sigma$ and the angle of oscillations
$\theta_{23}=\theta_{\mathrm{atm}}$ within $3\sigma$, we determined
the lowest limit of $10^4$~GeV for the mass of the heavy neutrino
singlet.

\section{Results for the case $\mathbf{n_R=2}$}
\label{3x2}

\subsection{General parameterization}

If we add two singlet fields $\nu_R$ to three left-handed neutrino
fields $\nu_L$, the radiative corrections give masses to all three
light neutrinos.
In the general case we parameterize
\begin{equation}
\Delta_i=\frac{\sqrt{2}}{v}\,
\left(
\begin{array}{c}
 m_{D_a} \vec{a}_i^{\top} \\
 m_{D_b}\vec{b}_i^{\top}
\end{array}
\right)
\end{equation}
with 12 complex parameters of $\vec{a}_i$ and $\vec{b}_i$, where
$i=1,2$. We further assume that the vectors are normalized,
$|\vec{a}_i|=|\vec{b}_i|=1$.
A specific set of
the vector $b$ values \refeq{specB} was used for this study.

Numerical evaluation of the model parameters and the masses of the
light neutrinos is performed in several steps. First, the neutrino
mass matrix for tree level is constructed. The lightest neutrino is
massless at tree level, $m_{l_1}^{(0)}=0$.
Taking $m_{l_1}^{\mathrm{in}}=0$, the masses of the other
two light neutrinos, $m_{l_2}^{\mathrm{in}}$ and $m_{l_3}^{\mathrm{in}}$, are
estimated from the experimental neutrino mass differences like for the
case $n_R=1$.
Entries of the heavy neutrino mass matrix,
$\hat{M}_R=\mathrm{diag}(m_{R_1},m_{R_2})$,
are input parameters. The eigenvalue equation in the block form is:
\begin{equation}\label{Mneutr2}
U_{\mathrm{tree}}^{T}M_{\nu}^{(0)}U_{\mathrm{tree}} =
U_{\mathrm{tree}}^{T}\left(
  \begin{array}{cc} 0_{3 \times 3} &
  m_{D_a} \vec{a}_1 \hspace{.2cm} m_{D_b} \vec{b}_1
\\
  \begin{array}{c}
    m_{D_a} \vec{a}^{\top}_1 \\ m_{D_b} \vec{b}^{\top}_1
  \end{array}
  &
  \hat{M}_R \end{array} \right)U_{\mathrm{tree}}
=\left( \begin{array}{cc} \hat{M}_l^{(0)} & 0 \\
  0 & \hat{M}_h^{(0)} \end{array} \right),
\end{equation}
where $\hat{M}_h^{(0)}=\mathrm{diag}(m_{h_1}^{(0)},m_{h_2}^{(0)})$
with $m_{h_1}^{(0)}\leq m_{h_2}^{(0)}$.
The values of $m_{D_a}$ and $m_{D_b}$
are evaluated through the seesaw equations,
relating $m_{D}$, $m_{R}$ on one side and $m_{l}$,
$m_{h}$ on the other side:
\begin{align}\label{seesaw2}
   m_{D_a}^2 =  m_{R_1}m_{l_2}^{\mathrm{in}}
          & \approx m_{h_1}^{(0)}m_{l_2}^{(0)} , \\
   m_{D_b}^2 =  m_{R_2}m_{l_3}^{\mathrm{in}}
          & \approx m_{h_2}^{(0)}m_{l_3}^{(0)} , \\
   m_{R_i}^2 \phantom{\ = \big(m_{h_i}^{\mathrm{in}}\big)^2}
            & \approx \bigl(m_{h_i}^{(0)}\bigr)^2,\ \ i=1,2.
\end{align}

The diagonalization matrix for tree level
$U_{\mathrm{tree}}=U_{\mathrm{egv}}^{\mathrm{tree}}\hat{U}_{\phi}(\phi_i)$
is composed of the eigenmatrix of $M_{\nu}^{(0)\dagger}M^{(0)}_{\nu}$
(denoted by $U_\mathrm{egv}^\mathrm{tree}$), and a diagonal phase
matrix $\hat{U}_{\phi}$. At the second step we evaluate one-loop
corrections to the neutrino mass matrix. Diagonalization of
$M_{\nu}^{(1)}$ is performed with a unitary matrix
$U_{\mathrm{loop}}=U_{\mathrm{egv}}^{\mathrm{loop}}\hat{U}_{\varphi}(\varphi_i)$,
where $U_{\mathrm{egv}}^{\mathrm{loop}}$ is the eigenmatrix of
$M^{(1)\dagger}_{\nu}M^{(1)}_{\nu}$, and $\hat{U}_{\varphi}$ is a
phase matrix. This procedure yields masses for all three light
neutrinos. If the calculated mass differences are compatible with the
experimental data on neutrino oscillations, the model parameter set
kept. Otherwise, another set of input parameters
$\{m_{R_{1,2}},\vec{a}_i,\vec{b}_i,m_{H_{2,3}^0}\}$ is generated.

The numerical analysis with the full set of parameters, constrained only
to the experimental mass differences of the light neutrinos,
suggests that heavy singlets should have
mass greater than 100~GeV~\cite{Jurciukonis:2012jz}.

\subsection{Reduced parameterization}

\def\BigColSep{\setlength{\arraycolsep}{1.5pt}}
\renewcommand{\arraystretch}{1.1}
\begin{table*}
\begin{center}
\begingroup\BigColSep
\begin{tabular}{|c|c|c|c|c|}
\hline \hline
\multirow{2}{*}{\parbox[t]{5em}{\centering2HDM\\ scenario}}
& \multicolumn{2}{|c|}{Normal hierarchy} & \multicolumn{2}{|c|}{Inverted hierarchy}\\ \cline{2-5}
 & $\Delta_{1} $ & $\Delta_{2}$ & $\Delta_{1}$ &$\Delta_{2}$\\
\hline
I & $\left( \begin{array}{ccc} a_{11}&a_{12}&a_{13}\\ 0&b_{12}&b_{13} \end{array} \right)$ & $\left( \begin{array}{ccc} 0&a_{22}&a_{23}\\ 0&b_{22}&b_{23} \end{array} \right)$ & $\left( \begin{array}{ccc} a_{11}&a_{12}&a_{13}\\0&b_{12}&b_{13} \end{array} \right)$ & $\left(\begin{array}{ccc} a_{21}&a_{22}&0\\ b_{21}&0&b_{23} \end{array} \right)$ \\
\hline
II & $\left( \begin{array}{ccc} a_{11}&a_{12}&a_{13}\\ 0&b_{12}&b_{13} \end{array} \right)$ & $\left( \begin{array}{ccc} a_{21}&a_{22}&0\\ 0&b_{22}&b_{23} \end{array} \right)$ & $\left( \begin{array}{ccc} a_{11}&a_{12}&a_{13}\\0&b_{12}&b_{13} \end{array} \right)$ & $\left(\begin{array}{ccc} 0&a_{22}&a_{23}\\ b_{21}&0&b_{23} \end{array} \right)$ \\
\hline
III$^{\mathrm{a,b}}$ & $\left( \begin{array}{ccc} a_{11}&a_{12}&0\\ 0&b_{12}&b_{13} \end{array} \right)$ & $\left( \begin{array}{ccc} 0&a_{22}&a_{23}\\ b_{21}&0&b_{23} \end{array} \right)$ & $\left( \begin{array}{ccc} a_{11}&a_{12}&a_{13}\\0&b_{12}&b_{13} \end{array} \right)$ & $\left(\begin{array}{ccc} a_{21}&a_{22}&0\\ b_{21}&0&b_{23} \end{array} \right)$ \\
\hline\hline
\end{tabular}
\endgroup
\end{center}
\caption{ Textures of Dirac matrices used for calculations of the
 light neutrino mass spectra for normal and inverted hierarchies in
 four different cases of the 2HDM vectors $b$, listed in
 Table~\ref{table1}.} \label{table2}
\end{table*}
\renewcommand{\arraystretch}{1.0}

Studying the influence of the randomly generated parameters we found that
a reduced number of parameters is sufficient to fulfil the
experimental criteria~\cite{Forero:2014bxa} of $\Delta m^2_\odot$, $\Delta
m^2_\mathrm{atm}$, $\theta_{12},$ $\theta_{13}$ and $\theta_{23}$.
We selected several ``textures'' of Dirac
matrices (i.e.\ the patterns of non-zero components), which allow the
most accurate agreement to the experimental data. The texture of the
matrix $\Delta_1$ has the largest impact to the results of oscillation
angles. There are 3 best versions of the $\Delta_1$ textures:
\begin{equation}
\left( \begin{array}{ccc} a_{11}&a_{12}&a_{13}\\ 0&b_{12}&b_{13} \end{array} \right),\
\left( \begin{array}{ccc}
 a_{11}&0&a_{13}\\ 0&b_{12}&b_{13} \end{array} \right),\
\mathrm{ and }\
\left( \begin{array}{ccc} a_{11}&a_{12}&0\\ 0&b_{12}&b_{13} \end{array} \right).
\end{equation}
The textures of the matrix $\Delta_2$ play a subdominant role for the
results. In our calculations we tailored the textures to the 2HDM
scenarios. As a result, the second texture of $\Delta_1$,
having $a_{12}=b_{11}=0$,
was not used. The applied textures of $\Delta_1$ and $\Delta_2$ are
listed in Table~\ref{table2}. All non-vanishing components can have
complex values.

Studies of textures with one or two zero entries in the models with
two heavy singlets have been reported in
refs.\ \cite{Ibarra:2003up,Guo:2006qa,Branco:2005jr,Rodejohann:2012jz,Lavoura:2013ysa,King:2013iva}.
Neither of those studies considered the SM extension by 2HDM.
We obtain experimentally
compatible neutrino oscillation values as in the mentioned references,
but our textures are tuned to provide the model parameters in best
agreement with the experimental data. It is worth mentioning that the
textures listed in Table~\ref{table2} may be a natural consequence of
a certain flavor symmetry, as discussed in
refs.~\cite{Raby:2003ay,Kuchimanchi:2002yu,Dutta:2003ps}.

The numerical evaluation of the model parameters and of the
light neutrino masses,
using the textures listed in Table~\ref{table2},
is done similarly to the general case.
Having picked a mass for the lightest neutrino as an input value
$m_{\mr{in}}\equiv m_{l_1}^{\mathrm{in}}$, the masses of the other two
light neutrinos, $m_{l_2}^{\mathrm{in}}$ and $m_{l_3}^{\mathrm{in}}$,
are estimated from the central values of experimental neutrino mass
differences. Entries of the heavy neutrino mass matrix,
$\hat{M}_R=\mathrm{diag}(m_{R_1},m_{R_2})$, are input parameters.  The
seesaw relations~\refeq{seesaw2} give the values of $m_{D_a}$ and
$m_{D_b}$. The
tree-level neutrino mass matrix $M_\nu^{(0)}$ is diagonalized, and
$U_{\mathrm{tree}}$ is obtained. Then one-loop corrections to the
neutrino mass matrix are calculated, and the parameters defining
$\Delta_2$ enter into the further evaluation. Diagonalization of the corrected
neutrino mass matrix $M_{\nu}^{(1)}$ yields masses for all three light
neutrinos. If the calculated mass differences are compatible with the
experimental data on oscillations, including the three neutrino mixing
angles, the model parameter set
is kept. Otherwise, another set of input parameters
$\{m_{\mr{in}},m_{R_{1,2}},\vec{a}_i,\vec{b}_i,\alpha_{ij},m_{H_{2,3}^0}\}$
is generated.

To increase the efficiency of random sampling, we chose the initial
mass $m_{\mr{in}}$ in the range of $16$--$20$~meV
according to the agreement of the
calculated results to the experimental values of the oscillation
angles. Unlike the case $n_R=1$, the oscillation angles are less
affected by the value of $m_{\mr{in}}$.
Calculation of radiative corrections is done for
the scenarios of the orthogonal complex vectors $b$ listed in
Table~\ref{table1}. The mass of the lightest neutral Higgs is fixed at
$m_{H^0_1}=125$~GeV but the masses of heavier neutral Higgses,
$m_{H^0_2}$ and $m_{H^0_3}$, are generated randomly in the range from
126 to 3000~GeV.
\begin{figure}[t]
\begin{center}
\includegraphics[scale=0.95]{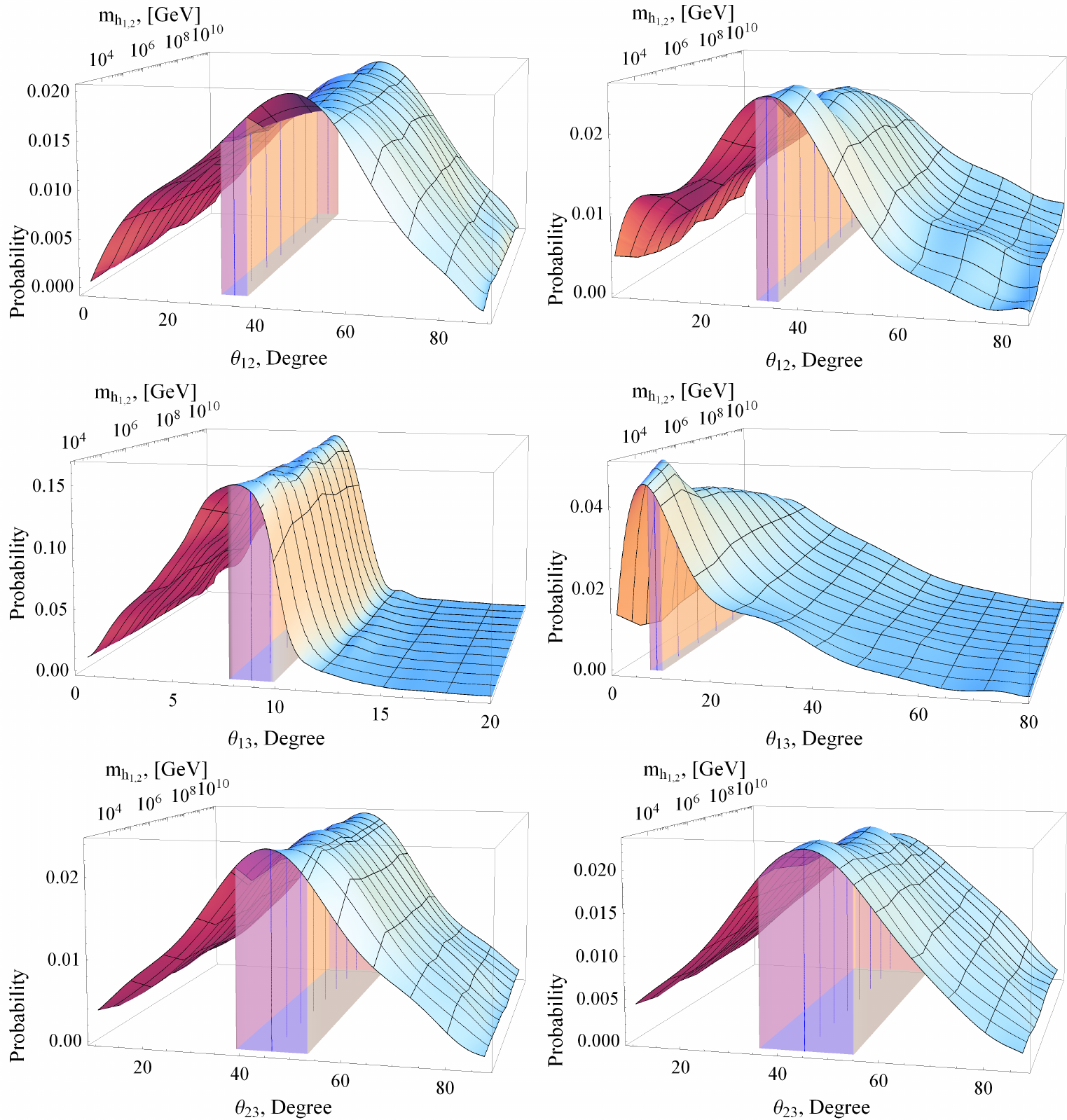}
\end{center}
\caption{(Color online) The 3D histograms of the oscillation angles
 for the scenario III in the case $n_R=2$. The plots on the left
 (right) correspond to the normal (inverted) hierarchy of the light
 neutrino masses. Filled boxes mark the applied experimental
 boundaries of $3\sigma$, blue vertical lines mark the experimental
 central values. }
\label{picture6}
\end{figure}

Figure~\ref{picture6} shows the distributions of the oscillation
angles for the scenario III using the best textures for $\Delta_1$ and
$\Delta_2$. Both normal and inverted neutrino mass hierarchies are
shown. In the case of the inverted hierarchy the peaks of the
distributions agree very well with the experimental bounds of all
oscillation angles.
In the case of the normal hierarchy, the most
probable value of $\theta_{12}$ is slightly different from the experimental
value. It is around $55$ degrees, instead of the expected
$33.4$ degrees. However, if we extract $\theta_{12}$ from $\sin^2(2\theta_{12})$ we will find two solutions in the first quadrant. Following
PDG~\cite{PDG2014}, the experimental boundaries are in the regions
$32.6$--$34.3$ degrees and $55.7$--$57.4$ degrees which agree with the peak of the atmospheric angle. In figure~\ref{picture6} the resulting values of $\theta_{13}$ are more localized as compared to the inverted hierarchy
case. The distributions of $\theta_{23}$ agree with the experimental
values in both hierarchies. The same tendency of the mixing angle
distributions is seen in all three scenarios of the 2HDM basis. It
should be noted that for the general parameterization case (where we
have 12 complex parameters) all peaks of the oscillation angle
distributions take values at approximately $45$ degrees.
\begin{figure}[t]
\begin{center}
\includegraphics[scale=0.95]{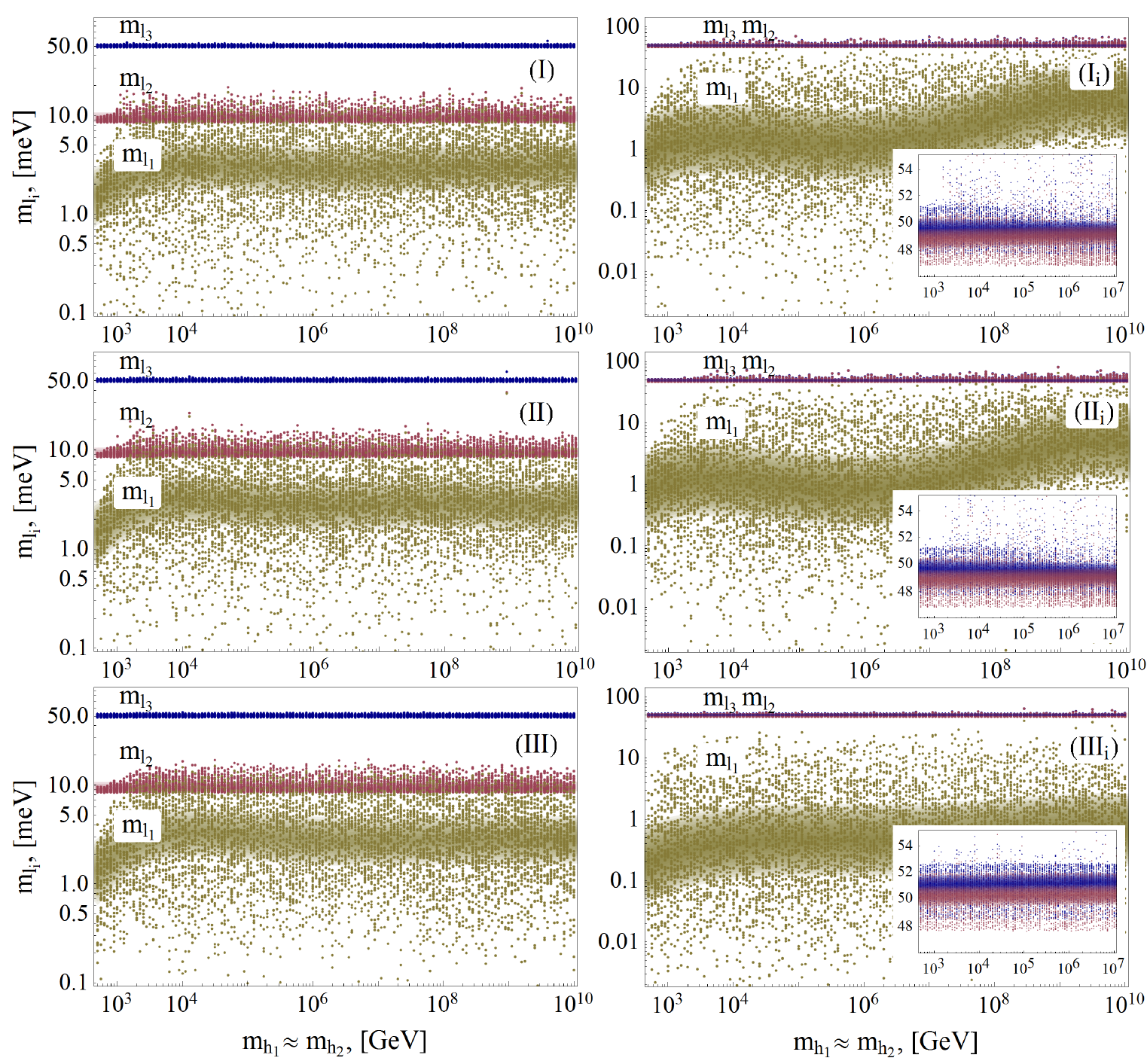}
\end{center}
\caption{(Color online) Masses $m_{l_i}$ of the light neutrinos as
 functions of the heaviest neutrino mass,
 $\max(m_{h_1},m_{h_2})$, for the scenarios of 2HDM in the case of
 $n_R=2$ (the scenarios IIIa and IIIb are
 very similar). The plots on the left represent the normal hierarchy, the
 plots on the right represent the inverted hierarchy of the light
 neutrinos. Wide bands indicate the area of the most
 frequent values of the scatter data. The nearly-degenerate masses
 $m_{l_2}$ and $m_{l_3}$ are shown separately in the lower right
 plots for the inverted hierarchy. }
\label{picture7}
\end{figure}

The neutrino mass spectrum is analyzed assuming that the masses of the
heavy neutrinos are nearly equal, $m_{h_1} \approx m_{h_2}$ (the ratio
$m_{h_1}/m_{h_2}$ is 0.999). This simplifies the analysis and does not
change the distributions of the light neutrino mass spectra in general
terms (we can compare figure~\ref{picture7} to figure~2 in
ref.~\cite{Jurciukonis:2012jz}). The masses of the light neutrinos
$m_{l_i}$ are estimated when the masses $m_{h_{1,2}}$ vary in the
range from $500$~GeV to $10^{10}$~GeV, see figure~\ref{picture7}. Both
normal and inverted hierarchies are shown. The dependency of the light
neutrino masses on the values of the heavy neutrino masses is similar
in all three scenarios. If normal hierarchy is assumed, the lightest
neutrino mass $m_{l_1}$ that is generated by the one-loop corrections
varies from $0.01$ to $20$~meV. The most frequent values lie
around $3$~meV, when the mass of the heavy neutrinos is
$m_{h_{1,2}}>3000$~GeV. The
mass $m_{l_2}$ varies from $8$ to $25$~meV with the most frequent
values at $10$~meV. The mass of the heaviest light neutrino $m_{l_3}$
is around $50$~meV. If the inverted hierarchy is assumed, the range of
$m_{l_1}$ values is wider and varies from $0.001$ to $40$~meV. The
most frequent values increase and take values of $0.5$--$4$~meV,
depending on the masses of the
heavy neutrinos. The values of $m_{l_2}$
and $m_{l_3}$ are nearly degenerate and vary from $47$ to
$55$~meV. The most frequent values are in the range of $48$--$51$~meV.
\begin{figure}[t]
\begin{center}
\includegraphics[scale=0.95]{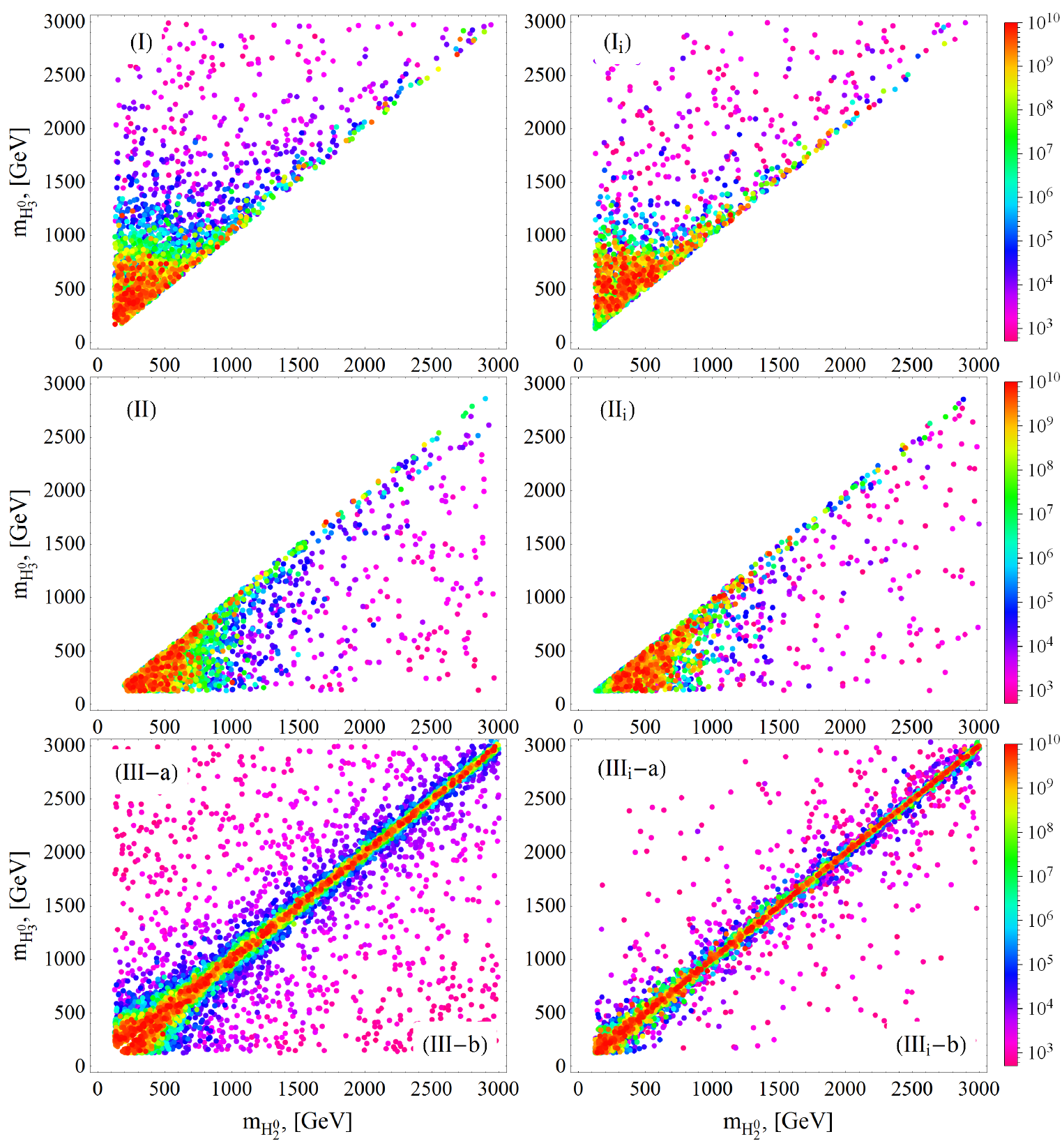}
\end{center}
\caption{(Color online) Values of the free parameters $m_{H^0_2}$ and
 $m_{H^0_3}$ as functions of the heaviest neutrino mass
 $\max(m_{h_1},m_{h_2})$ for the scenarios of 2HDM in the case
 $n_R=2$. The plots on the left represent the normal hierarchy, the
 plots on the right represent the inverted hierarchy of the light
 neutrinos. The scale of the $\max(m_{h_1},m_{h_2})$ values is shown
 on the right.}
\label{picture9}
\end{figure}

Figure~\ref{picture9} illustrates the allowed values of the Higgs
masses depending on the masses of the heavy singlets
$m_{h_{1,2}}$. Scenarios I and II are rather similar in
dependencies, namely, an increase of the heavy singlet mass leads to the
decrease of the Higgs masses. The mass of the second Higgs boson
tends to be different from the mass of the third Higgs boson.
The scenario III has different dependencies. The heavy Higgs masses
tend to be equal for large values of the heavy singlet masses, and
tend to be independent of it.

The values of the Higgs masses displayed in figure~\ref{picture9} also
satisfy the experimental restrictions \cite{PDG2014} of the oblique
parameters S, T, and U, as discussed in previous chapter. Using the
stronger constraint of $U=0$, about 40\%--65\% of the values shown in
figure~\ref{picture9} satisfy the experimental restrictions. In this
case the majority of $m_{H^0_{2,3}}$ values fall in the range of
126--1000~GeV.

The presented results are obtained using the tuned textures of
Table~\ref{table2}.
An alternative method to reduce the parameter
space of vectors $\vec{a}$ and $\vec{b}$ could be used. For example,
instead of setting an entire component ($a_{ij}$ or $b_{ij}$) to zero,
the parameter space could be limited restricting the values of these
vectors to real numbers.  A study of these textures will be reported
in the future.

The most reduced parameterization of Dirac matrices, that still allows
experimentally-compatible results of the light neutrino mass
differences, has only 4 independent real parameters:
\begin{align}
\Delta_1 = &\frac{\sqrt{2}}{v}\left(
    \begin{array}{ccc}
      m_{D_a} a_{1} & m_{D_a} \sqrt{1-a_{1}^2} & 0 \\
      0 & m_{D_b} b_{1} & m_{D_b}\sqrt{1-b_{1}^2}
    \end{array}
    \right),\\
\Delta_2 = &\frac{\sqrt{2}}{v}\left(
    \begin{array}{ccc}
      0 & m_{D_a} a_{2} & m_{D_a}\sqrt{1-a_{2}^2} \\
      m_{D_b} b_{2} & m_{D_b}\sqrt{1-b_{2}^2} & 0
    \end{array}
   \right) ,
\label{red5x5_2}
\end{align} 
with $|a_i|\leq1$ and $|b_i|\leq1$, $i=1,2$.
The neutrino oscillation angles evaluated in this strongly-reduced
parameterization do not have the most-probable values in the
experimentally determined range.

\section{Summary} \label{summary}

The seesaw mechanism is one of the most successful extensions of the
SM which explains neutrino masses. Finite corrections to the neutrino
mass matrix arise from one-loop diagrams mediated by a heavy
neutrino. In our model the Higgs sector is constructed with two Higgs
doublets and a CP-invariant Higgs potential which allows to
distinguish four conditions for vectors $b$ and thereby determine the
scenarios (see Table~\ref{table1}) for numerical calculations. The SM
Higgs mass is fixed to 125~GeV. By diagonalizing the neutrino mass matrix
we obtain light neutrino masses and derive their oscillation angles.
Sets of free parameters have been selected according to the
distributions of oscillation angles within the experimental boundaries, 
when also the masses of the light neutrinos give the measured 
neutrino mass differences.
In this paper we have studied two cases when one or two heavy
neutrinos are added to the three light neutrinos. We refer to those
cases as $n_R=1$ and $n_R=2$.

In the $n_R=1$ case with a general parametrization of the Dirac matrices
there are six free
complex parameters. The numerical analysis suggests a lower limit of 830~GeV
for the heavy singlet mass.
However, the large number of free
parameters makes it difficult to find the correlations among them. By reducing
the number of parameters we can study relations between them and the
dependency on the
heavy neutrino mass.  The minimal reduction of free parameters, namely,
$\vec{a}_1^{\top} = \bigl(0,a_{12},a_{13}\bigr)$ and
$\vec{a}_2^{\top}=\bigl(a_{21},a_{22},a_{23}\bigr)$, allows to estimate
all three oscillation angles $\theta_{12}$, $\theta_{13}$ and
$\theta_{23}$. The peaks of their distributions agree well with the
experimental bounds. We presented results of a strongly reduced
parameterization, $\vec{a}_1^{\top}=\bigl(0,0,1\bigr)$ and
$\vec{a}_2^{\top}=\bigl(0,n,e^{i\phi}\sqrt{1-n^2}\bigr)$.
According to the chosen minimal set of the free
parameters only the angle $\theta_{23}$ can be estimated. The calculated
masses of the light neutrinos decrease, when the heavy neutrino mass
increases. This dependency emerges from the relation of $m_{h}$ and
$\theta_{23}$. Tuning the initial value $m_{\mr{in}}$, and restricting
the light neutrino mass differences to the experimental central values
and the oscillation angle $\theta_{23}$ within $3\sigma$, we
determined for the mass of the heavy
neutrino singlet a lowest limit of $10^4$~GeV.
When the mass of the heavy singlet $m_h$ is
increasing in the scenarios I and II of the 2HDM model,
the allowed values of the non SM Higgs masses tend to decrease.
This tendency is absent in the scenario III, where the Higgs boson masses
$m_{H_{2,3}^0}$ get closer to each other, but stay different. 
We find a lower limit for
the allowed values of the heavier Higgs boson mass of about 500~GeV.
The values
of the free parameters depend weakly on the mass of the heavy singlet
$m_h$.

The general parametrization of the Dirac matrices in the $n_R=2$ case
has twelve complex
parameters. The numerical analysis shows that the heavy singlets should have
masses greater than $100$~GeV. However, the most probable values of the
neutrino oscillation angles
are not in the experimentally determined range. We selected several
textures of Dirac matrices, which allow the most accurate agreement to
the experimental data. The texture of the matrix $\Delta_1$ has the
largest impact on the values of the oscillation angles while the textures
of the matrix $\Delta_2$ play a sub-dominant role. The used
textures for normal and inverted neutrino mass hierarchies are listed in
Table~\ref{table2}.  The neutrino mass spectrum is analyzed assuming
that the masses of the heavy neutrinos are nearly equal, $m_{h_1}
\approx m_{h_2}$ (the ratio $m_{h_1}/m_{h_2}$ is 0.999). The masses of
the light neutrinos are estimated when the masses $m_{h_{1,2}}$ are
greater than $500$~GeV. 
The dependency of the light neutrino masses on
the values of the heavy neutrino masses is similar in all three
scenarios. An increase of the heavy singlet
mass leads to the decrease of the non SM Higgs masses in the scenarios
I and II.
This tendency is absent in the scenario
III, where the Higgs boson masses
$m_{H_{2,3}^0}$ tend to be equal as the masses of the heavy singlets increase.
The allowed values of $m_{H_{2,3}^0}$ can sample the entire range.
Due to the large number of free parameters it is
difficult to find correlations among them.

Our analysis has shown that the radiative corrections are quite
sizeable and play an important role. They should be taken into account
in the studies of the see-saw models.  The studied case with one heavy
singlet is a "toy" model because it is strongly restricted and does
not provide all physical quantities. For example, the mass of the
lightest neutrino is equal to zero, and we can evaluate only one
oscillation angle. However, using this model it is possible to make
some generalisations about the distributions of the Higgs masses and
the correlations between the free parameters for models with a larger
number of heavy singlets. The $n_R=2$ case allows the calculation of all
three masses of the light neutrinos with a reduced number of free
parameters. The finding of textures which allow the most accurate
agreement of the oscillation angles to the experimental data could
motivate some future models, for example, those based on the Abelian
family symmetry or another discrete symmetry.

\appendix

\section{Neutral Higgs mass eigenfields}
\label{Appendix-b-vectors}

Some features of formalism for the scalar sector of the
multi-Higgs-doublet SM is given in
ref.~\cite{Grimus:1989pu,Grimus:2002prd}. Here we discuss the
properties of the vectors $b$ and give expressions for their
calculation in the case of two Higgs doublets.

The physical neutral scalar mass eigenfields are expressed as
\begin{equation}\label{sumb-appendix}
\phi_{b_k}^0=\sqrt{2} \sum_{j=1}^{n_H} \mathrm{Re}(b_{kj}^* \phi_j^0)
= \frac{1}{\sqrt{2}} \sum_{j=1}^{n_{H}}
 \left( b^{*}_{kj} \phi^{0}_j +
   b_{kj} \phi^{0\,*}_j \right),
\end{equation}
which are characterized by $2 n_H$ unit vectors $b_k \in
\mathbbm{C}^{n_H}$ of dimensions $n_H \times 1$.
In the matrix-vector
notation, these eigenfields can be written as
$\phi_{b_k}^0=\sqrt{2}\,\mathrm{Re}(b_k^{\dagger} \phi^0)$.

The orthonormality equations for the vectors are
\begin{eqnarray}
&&\sum\limits_{j=1}^{n_H} \left(\mathrm{Re}(b_{kj})
 \mathrm{Re}(b_{k^{\prime}j}) + \mathrm{Im}(b_{kj})
 \mathrm{Im}(b_{k^{\prime}j})\right)
=
\sum\limits_{j=1}^{n_H} \mathrm{Re}(b_{kj}^* b_{k^{\prime}j})
  =\delta_{b_k b_{k^{\prime}}};
  \label{orthog1a}
\\
&& \sum\limits_{k=1}^{2n_H} \mathrm{Re}(b_{kj}) \mathrm{Re}(b_{k j^{\prime}})
  =\sum\limits_{k=1}^{2n_H} \mathrm{Im}(b_{kj}) \mathrm{Im}(b_{k j^{\prime}})
  =\delta_{jj'};
  \label{orthog1b}
\\
&& \sum\limits_{k=1}^{2n_H} \mathrm{Re}(b_{kj}) \mathrm{Im}(b_{k j^{\prime}})
 =\sum\limits_{k=1}^{2n_H} b_{kj} b_{k j^{\prime}}=0.
 \label{orthog1c}
\end{eqnarray}
The vectors $b_k$ and $b_{k^{\prime}}$ indicate two different states
$\phi_{b_k}^0$ and $\phi_{b_{k^{\prime}}}^0$, and indices $j$ and $j'$
indicate two different components of the vectors $b$.

The neutral Goldstone boson $G^0 = \phi^0_{G^0}$
corresponds to the vector $b_{G^0}$ with the components
$\left( b_{G^0} \right)_j = i v_j / v$ \cite{Grimus:1989pu,Grimus:2002prd},
where
$v = \left( \left| v_1 \right|^2 + \left| v_2 \right|^2
+ \cdots + \left| v_{n_H} \right|^2 \right)^{1/2}
= 2 m_W / g$.
In the case of only two Higgs doublets, and due to the rotation of the
Higgs fields to make the vacuum expectation value a feature of the
SM Higgs field, the vector $b_{G^0}$ equals
\begin{equation}\label{bG0}
b_{G^0}= \left(\begin{array}{c} i \\ 0 \end{array}\right).
\end{equation}

Physical Higgs fields $\phi^0_{b_k \neq G^0}$ must be orthogonal to
the Goldstone field
$G^0$
which follows from \refeq{orthog1a}. This leads to the condition
\begin{equation}
\sum_{j=1}^{n_H} \mathrm{Re} \left( -\frac{i v_j}{v} b^*_{kj} \right) =
 \frac{1}{v} \sum_{j=1}^{n_H} \mathrm{Im} \left( v_j b^*_{kj} \right) =
 \sum_{j=1}^{n_H} \mathrm{Re} \left( b_{G^0j} \, b^*_{kj} \right)=0.
\label{orthog2-append}
\end{equation}

To study the unit vectors $b$, introduced in eq.~\refeq{sumb-appendix}
(which is the same as eq.~\refeq{sumb} in the text) and corresponding to
the Higgs fields other than the Goldstone boson $G^0$, lets define
them in the following form:
\begin{equation}
b_1 = \left( \begin{array}{c} b_{11} \\ b_{12} \end{array} \right) ,
\qquad
b_2 = \left( \begin{array}{c} b_{21} \\ b_{22} \end{array} \right),
\qquad
b_3 = \left( \begin{array}{c} b_{31} \\ b_{32} \end{array} \right).
\label{bVectSetGeneral}
\end{equation}
From the orthogonality relations (\ref{orthog1a} - \ref{orthog1c}) and
due to the fixed value of $b_{G^0}$~\refeq{bG0} it is possible to write the
orthogonality equations for vector components in the following manner:
\begin{eqnarray}
&&
 b_{11}, b_{21}, b_{31} \in \mathbbm{R};
 \qquad
 b_{12}, b_{22}, b_{32} \in \mathbbm{C};
 \label{beq1a}
\\
&&
 b_{k1}^{2} + \left| b_{k2} \right|^2 = 1;
 \label{beq1b}
\\
&&
 b_{k1} b_{k^{\prime}1}
 +\mathrm{Re}\left( b_{k2}^{\star} b_{k^{\prime}2} \right)
 =0;
\\
&&
 \sum_{k=1}^{3} b^2_{k2} = \sum_{k=1}^{3} b_{k1} b_{k2} = 0;
\\
&&
 \sum_{k=1}^{3} b^{2}_{k1} =
 \sum_{k=1}^{3} \left[\mathrm{Re} \left(b_{k2} \right)\right]^2
 = \sum_{k=1}^{3} \left[\mathrm{Im} \left(b_{k2} \right)\right]^2 = 1.
 \label{beq1d}
\end{eqnarray}

By choosing $b_{31},$ $b_{21}$ and $\mathrm{Re} \left( b_{32}\right)$ as
input variables,
it is possible to express the other
components of the vectors $b$ by those variables by solving the
equations (\ref{beq1a} - \ref{beq1d}). Introducing three
sign-parameters $s_{32\mathrm{im}}$, $s_{11}$, and $s_{22}$ (they can take
values $\pm1$), we can write
\begin{align}
 \mathrm{Im} \left( b_{32}\right)=&
 s_{32\mathrm{im}} \sqrt{1- b_{31}^2-
   \left[\mathrm{Re} \left( b_{32}\right)\right]^2}\enspace;
 \label{beq2a}
\\
 b_{11}=& s_{11} \sqrt{1- b_{31}^2-b_{21}^2}\enspace;
 \label{beq2b}
\\
 b_{\mathrm{comb}}\equiv&
   \frac{b_{31} b_{21} \mathrm{Re} \left( b_{32}\right)
     +
     s_{22} \left | b_{11} \right|
       \left | \mathrm{Im} \left( b_{32}\right) \right|
   }
    {b_{31}^2 -1}\enspace
    ;
\\
 p_{22} \equiv& \left\{
\begin{aligned}
  - \mathrm{Sg}(b_{31}) \mathrm{Sg}(b_{21})
    \mathrm{Sg}(\mathrm{Im} \left( b_{32}\right)),&\
   \mathrm{if}\
    \left |\mathrm{Re} \left( b_{32}\right) \right|
       \leqslant \sqrt{\frac{b_{31}^2 b_{21}^2}{1-b_{21}^2}}\ ,
       \\
   s_{22}\mathrm{Sg}(\mathrm{Re}
    \left(b_{32}\right)) \mathrm{Sg}(\mathrm{Im} \left(
    b_{32}\right)),&\
     \mathrm{otherwise}\ ;
\end{aligned}
\right.
\\
 b_{22} =& b_{\mathrm{comb}} + i p_{22}
   \sqrt{1- b_{21}^2- b^2_{\mathrm{comb}}}\ ,
\label{beq2d}
\\
 b_{12}=& -\frac{1}{b_{11}}
 \left( b_{31} b_{32} + b_{21} b_{22} \right) .
 \label{beq2c}
\end{align}
We introduced two intermediate parameters $b_{\mathrm{comb}}$ and
$p_{22}$, and $\mathrm{Sg}(x)$ is the sign function
\begin{equation}
\mathrm{Sg}(x) =
\left\{ \begin{array}{c} -1,\hspace{0.2cm} x<0, \\ \phantom{-}
 1, \hspace{0.2cm} x \geqslant 0 . \end{array}\right.
\end{equation}
It is worth mentioning that the solutions for the parameter values,
given by the equations (\ref{beq2a} - \ref{beq2d}), were obtained
assuming $b_{21}, b_{31} \neq \pm 1$. According to the orthogonality
relations (\ref{beq1a} - \ref{beq1d}) the free scale parameters vary
in the following ranges: $|b_{31}|<1$, $|b_{21}|<\sqrt{1-b_{31}^2}$, and
$|\mathrm{Re}(b_{32})|\leq\sqrt{1-b_{31}^2}$. The
extreme values of $\pm1$ for the parameters $b_{21}$ and $b_{31}$
could be obtained by
the index permutation of the vectors $b_k$ (for example,
$b_{21}=1$ can be obtained by
swapping the values of $b_{11}=1$ and $b_{12}$ with those of
$b_{21}$ and $b_{22}$).

Equations (\ref{beq2a} - \ref{beq2d}) give 8 different solutions for
the vectors $b$, corresponding to two possible values of the
sign-parameters $s_x$ ($x=32\mathrm{im}$, $11$, and $22$). However,
due to the structure of the one-loop corrections
(\ref{corrections}), only 4 different solutions of those equations are
important, since the sign of $b_{11}$ (i.e.\ the
value of $s_{11}$) does not change the values of the light neutrino
masses.

The expressions of eqs.~(\ref{beq2a} - \ref{beq2d}) are significantly
simpler, if some input parameters are equal to zero. This can lead to
further simplification after introducing trigonometric functions.
Let us study the case, when $\mathrm{Re} \left( b_{32}\right) = 0$.
Defining $b_{31} = \sin(\alpha_{13})$,
$b_{21} = \sin(\alpha_{12})\cos(\alpha_{13})$, and taking
$s_{32\mathrm{im}}=s_{11}=s_{22}=1$,
we obtain the following parametric values of the vectors $b$:
\begin{equation}
b_{G^0} = \left( \begin{array}{c} i \\ 0 \end{array} \right),\
b_1 = \left( \begin{array}{c}
   \mr c_{12} \mr c_{13} \\ -\mr s_{12} - i \mr c_{12} \mr s_{13}
   \end{array} \right) ,\
b_2 = \left( \begin{array}{c}
    \mr s_{12} \mr c_{13} \\ \mr c_{12} - i \mr s_{12} \mr s_{13}
    \end{array} \right),\
b_3 = \left( \begin{array}{c}
    \mr s_{13} \\ i \mr c_{13}
   \end{array} \right),
\label{bVectSetTh-appendix}
\end{equation}
where $\mr c_{ij} \equiv \cos(\alpha_{ij})$ and $\mr s_{ij} \equiv
\sin(\alpha_{ij})$.

\section{Neutrino oscillation angles}
\label{Appendix-oscillation-angles}

Neutrino oscillation angles are introduced using the tree-level
neutrino mass diagonalization matrix $U_{\mathrm{loop}}$ and
factorizing it to contain the ordinary Pontecorvo-Maki-Nakagawa-Sakata
(PMNS) neutrino mixing matrix. We introduce the formalism by
discussing the $3\times3$ neutrino mixing case, where the
relationships are simpler. Then we discuss the cases that are used in
current paper, namely, $(3+1)\times(3+1)$ and $(3+2)\times(3+2)$
neutrino mixing.

The simplest case ($3\times3$) considers only the SM neutrinos. It is
discussed in ref.~\cite{Dziewit:2011pd} in a slightly different
notation of the matrix elements. Factorization of the rotation matrix
with the PMNS matrix included explicitly in the case $3+3$ is
discussed in ref.~\cite{Xing:2011ur}. Here we give formulas
for the intermediate cases.

The neutrino masses and the mixing angles are predicted from a given
neutrino mass matrix (the ``top-down'' method). Exact analytical
expressions for the mixing angles, Dirac and Majorana phases and
formulas for the non-physical phases can be given for the 3- and
4-dimensional cases. Only numerical solutions are possible in the
case of 2 additional neutrinos (the 5-dimensional case).

\bigskip
\noindent {\bf The 3-dimensional case}

First we parameterize the neutrino diagonalisation matrix by including
the PMNS mixing matrix for the $3\times3$ mixing.
The neutrino mass matrix can be diagonalised
by a unitary transformation $U$, obtained
by the singular value decomposition method, see eq.~\refeq{Mtotal}.
Lets denote the
matrix elements in the following way:
\begin{equation}
\label{apAU}
U^{\mathrm (3x3)}=
\left(
\begin{array}{ccc}
 x_1 & x_2 & x_3
  \\
 y_1 & y_2 & y_3
  \\
 z_1 & z_2 & z_3
\end{array}
\right).
\end{equation}
This matrix could be factorized into three terms
\begin{equation}
\label{apA2}
U^{\mathrm (3x3)} =
  \hat{U}^{(3)}_{\phi} \cdot U_{\mr{PMNS}} \cdot \hat{U}^{(3)}_{\kappa},
\end{equation}
where $U_{\mr{PMNS}}$ is the standard PMNS mixing matrix for Dirac neutrinos:
\begin{eqnarray}
U_{\mr{PMNS}}&=&\left( \begin{array}{ccc}
1 & 0 & 0 \\
0 & \mr{c}_{23} & \mr{s}_{23} \\
0 & -\mr{s}_{23} & \mr{c}_{23}
\end{array} \right) \cdot
\left( \begin{array}{ccc}
\mr{c}_{13} & 0 & \hat{\mr{s}}_{13}^* \\
0 & 1 & 0 \\
-\hat{\mr{s}}_{13} & 0 & \mr{c}_{13}
\end{array} \right) \cdot
\left( \begin{array}{ccc}
\mr{c}_{12} & \mr{s}_{12} & 0 \\
-\mr{s}_{12} & \mr{c}_{12} & 0 \\
0 & 0 & 1
\end{array} \right) \notag \\
&=& \left(
\begin{array}{ccc}
 \mr{c}_{12} \mr{c}_{13} & \mr{c}_{13} \mr{s}_{12} & \hat{\mr{s}}_{13}^* \\
 -\mr{c}_{23} \mr{s}_{12}-\mr{c}_{12} \hat{\mr{s}}_{13} \mr{s}_{23} & \mr{c}_{12} \mr{c}_{23}- \mr{s}_{12} \hat{\mr{s}}_{13} \mr{s}_{23} &
  \mr{c}_{13} \mr{s}_{23} \\
\mr{s}_{12} \mr{s}_{23}-\mr{c}_{12} \mr{c}_{23}\hat{\mr{s}}_{13} & -\mr{c}_{23} \mr{s}_{12} \hat{\mr{s}}_{13}-\mr{c}_{12} \mr{s}_{23} &
\mr{c}_{13} \mr{c}_{23}
\end{array}
\right) .
\end{eqnarray}
We used abreviations $\mr{c}_{ij} \equiv \cos\theta_{ij}$ and
$\hat{\mr{s}}_{ij} \equiv e^{i \delta_{ij}}\sin\theta_{ij}$, where
$\theta_{ij}$ and $\delta_{ij}$ are the rotation angle and the phase
angle, respectively.

The two diagonal phase matrices are defined as
\begin{eqnarray}
\hat{U}^{(3)}_{\phi} &=& \left( \begin{array}{ccc}
e^{i \phi_1} & 0 & 0 \\
0 & e^{i \phi_2} & 0 \\
0 & 0 & e^{i \phi_3}
\end{array} \right)\ ,
\\
\hat{U}^{(3)}_{\kappa} &=&\left( \begin{array}{ccc}
1 & 0 & 0 \\
0 & e^{i \kappa_1/2} & 0 \\
0 & 0 & e^{i \kappa_2/2}
\end{array} \right)\ .
\end{eqnarray}

There are 9 parameters: 3~mixing angles $(\theta_{12}$,
$\theta_{13}$, $\theta_{23})$; 1~Dirac phase $\delta_{13}$;
2~Majorana phases $\kappa_1$ and $\kappa_2$; and the matrix
$\hat{U}^{(3)}_{\phi}$ containing 3~non-physical and unmeasurable
phases $\phi_i$ ($i=1,2,3$).

Comparing eqs.~(\ref{apAU}) and (\ref{apA2}) we can find the relations
between the elements of the rotation matrix in a general form and in the
factorized form:
\begin{equation} \label{theta3x3}
\theta_{13}=\arcsin\left(\left|x_3\right|\right), \hspace{.2cm} \theta_{23}=\arcsin\left(\frac{\left|y_3\right|}{\sqrt{1-\left|x_3\right|^2}}\right), \hspace{.2cm} \theta_{12}=\arcsin\left(\frac{\left|x_2\right|}{\sqrt{1-\left|x_3\right|^2}}\right),
\end{equation}
\begin{equation}
\frac{\kappa_1}{2}=\mr{arg}(x_2)-\mr{arg}(x_1), \quad \frac{\kappa_2}{2}= \mr{arg}(x_3)-\mr{arg}(x_1) +\delta_{13},
\end{equation}
\begin{equation}
\phi_1=\mr{arg}(x_1), \quad \phi_2=\mr{arg}(x_1)-\mr{arg}(x_3)+\mr{arg}(y_3)-\delta_{13},
\end{equation}
\begin{equation}
\phi_3=\mr{arg}(x_1)-\mr{arg}(x_3)+\mr{arg}(z_3)-\delta_{13},
\end{equation}
\begin{equation}
\delta_{13}=\mr{arg}(x_2)-\mr{arg}(x_3)+\mr{arg}(y_3)+
i \ln \left(\frac{y_2 \left(1-|x_3|^2\right)+x_2\, x_3^*\, y_3}{|x_1||z_3|}\right).
\end{equation}
These relations are obtained comparing eq.~(\ref{apA2}) with the following
matrix elements from eq.~(\ref{apAU}), forming the upper-triangular matrix:
$x_1$, $x_2$, $x_3$, $y_2$, $y_3$,
and $z_3$. Other (identical) solutions are possible, using the matrix elements
$y_1$, $z_1$, and $z_2$.

\bigskip
\noindent {\bf 4-dimensional case}

If there is one additional neutrino, decomposition of the
neutrino mass diagonalization matrix into factors including the PMNS
neutrino mixing matrix is more complicated. Lets define
2-dimensional rotation matrices in the 4-dimensional complex space,
similarly to ref.~\cite{Xing:2011ur},
\begin{equation*}
R^{(4)}_{12} =\left( \begin{array}{cccc} \mr{c}_{12} & \mr{s}_{12} & 0 & 0 \\
-\mr{s}_{12} & \mr{c}_{12} & 0 & 0 \\ 0 & 0 & 1 & 0 \\ 0 & 0 & 0 & 1 \end{array} \right), \qquad
R^{(4)}_{13} =\left( \begin{array}{cccc} \mr{c}_{13} & 0 & \hat{\mr{s}}_{13}^* & 0 \\ 0 & 1 & 0 & 0 \\
-\hat{\mr{s}}_{13} & 0 & \mr{c}_{13} & 0 \\ 0 & 0 & 0 & 1 \end{array} \right),
\end{equation*}
\begin{equation*}
R^{(4)}_{23} =\left( \begin{array}{cccc} 1 & 0 & 0 & 0 \\
0 & \mr{c}_{23} & \mr{s}_{23} & 0 \\ 0 & -\mr{s}_{23} & \mr{c}_{23} & 0 \\ 0 & 0 & 0 & 1 \end{array} \right), \qquad
R^{(4)}_{14} =\left( \begin{array}{cccc} \mr{c}_{14} & 0 & 0 &\hat{\mr{s}}_{14}^* \\
0 & 1 & 0 & 0 \\ 0 & 0 & 1 & 0 \\ -\hat{\mr{s}}_{14} & 0 & 0 & \mr{c}_{14} \end{array} \right),
\end{equation*}
\begin{equation}\label{apB1}
R^{(4)}_{24} =\left( \begin{array}{cccc} 1 & 0 & 0 & 0 \\
0 & \mr{c}_{24} & 0 & \hat{\mr{s}}_{24}^* \\ 0 & 0 & 1 & 0 \\ 0 & -\hat{\mr{s}}_{24} & 0 & \mr{c}_{24} \end{array} \right), \qquad
R^{(4)}_{34} =\left( \begin{array}{cccc} 1 & 0 & 0 & 0 \\
0 & 1 & 0 & 0 \\ 0 & 0 & \mr{c}_{34} & \hat{\mr{s}}_{34}^* \\ 0 & 0 & -\hat{\mr{s}}_{34} & \mr{c}_{34} \end{array} \right),
\end{equation}
and phase matrices:
$\hat{U}^{(4)}_{\phi} =
\mr{diag}\left( e^{i \phi_1},e^{i \phi_2},e^{i \phi_3},e^{i \phi_4}\right)$
and
$\hat{U}^{(4)}_{\kappa} = \mr{diag}
  \left(1,e^{i \kappa_1 /2},e^{i \kappa_2/2},1 \right)$.
Note that a shorter notation can be used to define the elements of
the rotation matrices:
\begin{equation}
\big[R^{(4)}_{jk}\big]_{a}^{\,\,b} =
\delta_{a}^{\,b} 
+ ( \mr{c}_{jk} - 1 ) 
  ( \delta_{a}^{\,j} \delta_{j}^{\,b} 
  + \delta_{a}^{\,k} \delta_{k}^{\,b} )
+ \hat{\mr{s}}_{jk}^{*} \delta_{a}^{\,j} \delta_{k}^{\,b} 
- \hat{\mr{s}}_{jk} \delta_{a}^{\,k} \delta_{j}^{\,b}\ ,
\end{equation}
where $\delta_{a}^{\,b}$ equals $1$, when $a=b$, or $0$, otherwise. This
notation is not restricted to the 4-dimensional case.

The unitary matrix $U^{\rm (4x4)}$ is
parameterized by
\begin{equation}\label{apB3}
U^{\mr{(4x4)}}=
\hat{U}^{(4)}_{\phi}\cdot
\left(R^{(4)}_{34} R^{(4)}_{24} R^{(4)}_{14}\right)\cdot
\left(R^{(4)}_{23} R^{(4)}_{13} R^{(4)}_{12}\right)\cdot
\hat{U}^{(4)}_{\kappa} ,
\end{equation}
with the PMNS matrix defined by a product of
three rotation matrices:
\begin{equation}\label{apB4} \left(
\begin{array}{cc} U_{\rm PMNS} & \mathbf{0} \\
\mathbf{0} & 1 \end{array} \right) =
\left(R^{(4)}_{23} R^{(4)}_{13} R^{(4)}_{12}\right).
\end{equation}
There are 16 parameters in this case, namely: 6 mixing angles
$(\theta_{12},\theta_{13},\theta_{23},
\theta_{14},\theta_{24},\theta_{34})$; 1~Dirac phase $\delta_{13}$;
2~Majorana phases $\kappa_1$ and $\kappa_2$; 3~phases
$\delta_{14},\delta_{24},\delta_{34}$; and the matrix
$\hat{U}^{(4)}_{\phi}$ containing 4~phases $\phi_i$ ($i=1,2,3,4$).

For our model with $n_R=1$ we calculate the diagonalization matrix
$U_{\mathrm{loop}}$ numerically. Defining its elements as
\begin{equation}\label{apBU}
U^{\rm (4x4)}=
\left(
\begin{array}{cccc}
 x_1 & x_2 & x_3 & x_4
  \\
 y_1 & y_2 & y_3 & y_4
  \\
 z_1 & z_2 & z_3 & z_4
  \\
 t_1 & t_2 & t_3 & t_4
\end{array}
\right)
\end{equation}
and comparing to eq.~(\ref{apB3}) we find the relations:
\begin{eqnarray}\label{apB5}
&& \theta_{12}= \arcsin \left(\frac{|x_2|}{\sqrt{b}}\right), \quad \theta_{13}= \arcsin \left(\frac{|x_3|}{\sqrt{a}}\right), \quad
\theta_{23}= \arcsin \left(\frac{|d|}{\sqrt{b}\sqrt{c}}\right), \\
&& \theta_{14}= \arcsin \left(|x_4|\right), \quad
\theta_{24}= \arcsin \left(\frac{|y_4|}{\sqrt{a}}\right), \quad
\theta_{34}= \arcsin \left(\frac{|z_4|}{\sqrt{c}}\right),
\end{eqnarray}
\begin{equation}\label{apB8}
\phi_1 = \arg(x_1), \qquad \phi_4=\arg(t_4),
\end{equation}
\begin{align}\label{apB6}
\phi_2 =& \arg(x_1)-\arg(x_2) \notag \\ 
&-i \ln \left(
\frac{a\, b\, y_2 + x_2\, x_4^*\, y_4 \sqrt{b} \sqrt{a-|x_3|^2}+
    d\, |a|\, x_2\, x^*_3 \sqrt{a-|y_4|^2}
   \big/ \left(a \sqrt{c}\right)}
 {\sqrt{a}\,
     \sqrt{b\,c-|d|^2}\,\sqrt{b-|x_2|^2} }
\right),
\end{align}
\begin{align}\label{apB7}
\phi_3 = \phi_2 +i \ln \left(
\frac{d\, |a|\sqrt{a}\sqrt{b\,c-|d|^2}\, \sqrt{a-|x_3|^2} \sqrt{c-|z_4|^2}
   \big/ |d|}
 {a^2\sqrt{b}\, c\, z_3+
    a \sqrt{b}\sqrt{c}\,x_3\, x^*_4\, z_4 \sqrt{a-|y_4|^2}+
    d \,|a|\, y^*_4\, z_4 \sqrt{a-|x_3|^2} }
\right),
\end{align}
\begin{eqnarray}\label{apB9}
&& \delta_{13}=\arg(x_1)-\arg(x_3)-\phi_2-i\ln \left( \frac{d\, |a|}{a\, |d|} \right),   \\
&& \delta_{14}=\phi_1-\arg(x_4), \quad
\delta_{24}=\phi_2-\arg(y_4), \quad
\delta_{34}=\arg(z_4)-\phi_3, \\
&& \frac{\kappa_1}{2}=\arg(x_2)-\phi_1, \quad \frac{\kappa_2}{2}=-\phi_2-i\ln \left( \frac{d\, |a|}{a\, |d|} \right),
\end{eqnarray}
where:
\begin{eqnarray}
&& a=1-|x_4|^2, \qquad b=1-|x_3|^2-|x_4|^2, \notag \\
&& c=1-|x_4|^2-|y_4|^2, \qquad d=a\, y_3 + x_3\, x^*_4\, y_4.
\end{eqnarray}
Because of the discontinuous nature of the square root function in the
complex plane, $\sqrt{x y} \neq \sqrt{x} \sqrt{y}$ in general. 
Therefore a simplification of the above expressions is limited.

Due to its origin and the relations between the elements, the
expressions do not contain all entries of the rotation matrix
$U^{(4\times4)}$, defined in eq.~\refeq{apBU}. These relations
are obtained comparing eq.~(\ref{apB3}) with the following matrix
elements from eq.~(\ref{apBU}), forming the upper-triangular matrix:
$x_1$, $x_2$, $x_3$, $x_4$, $y_2$, $y_3$, $y_4$, $z_3$, $z_4$,
and $t_4$. Other (identical) solutions are possible using the matrix
elements $z_1$, $z_2$, $t_1$, $t_2$, and $t_3$.

\bigskip
\noindent {\bf 5-dimensional case}

To introduce factorization containing the PMNS neutrino mixing matrix
in the $3+2$ case, we first
define the rotation matrices in the 5-dimensional complex
space, similarly to ref.~\cite{Xing:2011ur}
\begin{equation*}
R^{(5)}_{12} =\left( \begin{array}{ccccc} \mr{c}_{12} & \mr{s}_{12} & 0 & 0 & 0 \\
-\mr{s}_{12} & \mr{c}_{12} & 0 & 0 & 0 \\ 0 & 0 & 1 & 0 & 0 \\ 0 & 0 & 0 & 1 & 0 \\ 0 & 0 & 0 & 0 & 1 \end{array} \right), \qquad
R^{(5)}_{13} =\left( \begin{array}{ccccc} \mr{c}_{13} & 0 & \hat{\mr{s}}_{13}^* & 0 & 0 \\ 0 & 1 & 0 & 0 & 0 \\
-\hat{\mr{s}}_{13} & 0 & \mr{c}_{13} & 0 & 0 \\ 0 & 0 & 0 & 1 & 0 \\ 0 & 0 & 0 & 0 & 1 \end{array} \right),
\end{equation*}
\begin{equation*}
R^{(5)}_{23} =\left( \begin{array}{ccccc} 1 & 0 & 0 & 0 & 0 \\
0 & \mr{c}_{23} & \mr{s}_{23} & 0 & 0 \\ 0 & -\mr{s}_{23} & \mr{c}_{23} & 0 & 0 \\ 0 & 0 & 0 & 1 & 0 \\ 0 & 0 & 0 & 0 & 1 \end{array} \right), \qquad
R^{(5)}_{14} =\left( \begin{array}{ccccc} \mr{c}_{14} & 0 & 0 & \hat{\mr{s}}_{14}^* & 0 \\
0 & 1 & 0 & 0 & 0 \\ 0 & 0 & 1 & 0 & 0 \\ -\hat{\mr{s}}_{14} & 0 & 0 & \mr{c}_{14} & 0 \\ 0 & 0 & 0 & 0 & 1 \end{array} \right),
\end{equation*}
\begin{equation*}
R^{(5)}_{24} =\left( \begin{array}{ccccc} 1 & 0 & 0 & 0 & 0 \\
0 & \mr{c}_{24} & 0 & \hat{\mr{s}}_{24}^* & 0 \\ 0 & 0 & 1 & 0 & 0 \\ 0 & -\hat{\mr{s}}_{24} & 0 & \mr{c}_{24} & 0 \\0 & 0 & 0 & 0 & 1 \end{array} \right), \qquad
R^{(5)}_{34} =\left( \begin{array}{ccccc} 1 & 0 & 0 & 0 & 0 \\
0 & 1 & 0 & 0 & 0 \\ 0 & 0 & \mr{c}_{34} & \hat{\mr{s}}_{34}^* & 0 \\ 0 & 0 & -\hat{\mr{s}}_{34} & \mr{c}_{34} & 0 \\ 0 & 0 & 0 & 0 & 1 \end{array} \right),
\end{equation*}
\begin{equation*}
R^{(5)}_{15} =\left( \begin{array}{ccccc} \mr{c}_{15} & 0 & 0 & 0 & \hat{\mr{s}}_{15}^* \\
0 & 1 & 0 & 0 & 0 \\ 0 & 0 & 1 & 0 & 0 \\ 0 & 0 & 0 & 1 & 0 \\ -\hat{\mr{s}}_{15} & 0 & 0 & 0 & \mr{c}_{15} \end{array} \right), \qquad
R^{(5)}_{25} =\left( \begin{array}{ccccc} 1 & 0 & 0 & 0 & 0 \\ 0 & \mr{c}_{25} & 0 & 0 & \hat{\mr{s}}_{25}^* \\ 0 & 0 & 1 & 0 & 0 \\ 0 & 0 & 0 & 1 & 0 \\ 0 & -\hat{\mr{s}}_{25} & 0 & 0 & \mr{c}_{25}\end{array} \right),
\end{equation*}
\begin{equation*}\label{apC1}
R^{(5)}_{35} =\left( \begin{array}{ccccc} 1 & 0 & 0 & 0 & 0 \\ 0 & 1 & 0 & 0 & 0 \\ 0 & 0 & \mr{c}_{35} & 0 & \hat{\mr{s}}_{35}^* \\
 0 & 0 & 0 & 1 & 0 \\ 0 & 0 & -\hat{\mr{s}}_{35} & 0 & \mr{c}_{35} \end{array} \right), \qquad
R^{(5)}_{45} =\left( \begin{array}{ccccc} 1 & 0 & 0 & 0 & 0 \\ 0 & 1 & 0 & 0 & 0 \\ 0 & 0 & 1 & 0 & 0 \\ 0 & 0 & 0 & \mr{c}_{45} & \hat{\mr{s}}_{45}^* \\
 0 & 0 & 0 & -\hat{\mr{s}}_{45} & \mr{c}_{45} \end{array} \right),
\end{equation*}
and the phase matrices:
\begin{align}
\hat{U}^{(5)}_{\phi} =& \mr{diag}\left(
 e^{i \phi_1},e^{i \phi_2},e^{i \phi_3},e^{i \phi_4},e^{i \phi_5}
\right), \\
\hat{U}^{(5)}_{\kappa} =& \mr{diag}\left(
1,e^{i \kappa_1/2},e^{i \kappa_2 /2},1,1 \right).
\end{align}
 The unitary matrix $U^{\rm (5x5)}$ is parameterized by
\begin{equation}\label{apC3}
U^{\mr{(5x5)}}=U^{(5)}_{\phi} \cdot
\left(R^{(5)}_{45} R^{(5)}_{35} R^{(5)}_{25} R^{(5)}_{15} \right)
\cdot
\left(R^{(5)}_{34} R^{(5)}_{24} R^{(5)}_{14}\right)
\cdot
\left(R^{(5)}_{23} R^{(5)}_{13} R^{(5)}_{12}\right)
\cdot U^{(5)}_{\kappa}
\end{equation}
with the inclusion of the PMNS matrix
\begin{equation}\label{apC4} \left(
\begin{array}{cc} U_{\rm PMNS} & \mathbf{0} \\
\mathbf{0} & \mathbf{1} \end{array} \right) = \left(R^{(5)}_{23} R^{(5)}_{13} R^{(5)}_{12}\right).
\end{equation}

There are 25 parameters in the 5-dimensional case: 10 mixing angles
$(\theta_{12},\theta_{13},\theta_{23},$
$\theta_{14},\theta_{24},\theta_{34},\theta_{15},$
$\theta_{25},\theta_{35},\theta_{45})$;
1 Dirac phase $\delta_{13}$; 2 Majorana phases $\kappa_1$ and
$\kappa_2$;
7 phases
$\delta_{14},\delta_{24},\delta_{34},\delta_{15},$
$\delta_{25},\delta_{35},\delta_{45}$; and
the matrix $\hat{U}^{(5)}_{\phi}$ containing 5
phases~$\phi_i,~i=1,\dots,5$. Only numerical solutions
for the parameters are possible.

A simplification is possible in our analysis. According
to the structure of the diagonalisation matrix $U_{\mr{loop}}$ in the
4- or 5-dimensional cases, the $3\times3$ sub-matrix in the top-left
corner is dominant. This sub-matrix corresponds to the matrix $U_{\rm PMNS}$.
In the numerical calculation, it suffices to use the expressions
\ref{theta3x3} in order to estimate the oscillation angles
$\theta_{12},\theta_{13}$ and $\theta_{23}$,
in the
cases of $n_R=1$ and $n_R=2$. There is a numerical precision
difference between the approximated angle
values and the values obtained using the expressions (\ref{apB5}) in
the 4-dimensional case or the numerical solutions in the 5-dimensional
case. The
approximation speeds up the calculations significantly.


\acknowledgments

The authors thank the Lithuanian Academy of Sciences for the
support (the project DaFi2015).
Special thanks to Luis Lavoura for valuable discussions and suggestions.



\bibliographystyle{JHEP}
\bibliography{neutrinoPaper_bib}

\end{document}